\documentclass[%
reprint,
amsmath,amssymb,
prb,
]{revtex4-2}

\usepackage{lipsum, babel}
\usepackage{xcolor}
\usepackage[utf8]{inputenc}
\usepackage{longtable}
\usepackage{CJK}
\usepackage{graphicx}
\usepackage{dcolumn}
\usepackage{bm}
\usepackage{booktabs}
\usepackage{soul}

\begin{document}
\begin{CJK*}{UTF8}{gbsn}
\preprint{APS/123-QED}

\title{{\it Ab initio} constraints on silica melting to  500 GPa}

\author{Ming Geng (耿明) $^{1,2}$}

\author{Chris E. Mohn$^{1,2,3}$}
\email{chrism@kjemi.uio.no}
\affiliation{
$^{1}$ Centre for Earth Evolution and Dynamics (CEED), University of Oslo, N-0315 Oslo, Norway
}
\affiliation{
$^{2}$ Centre for Planetary Habitability (PHAB), University of Oslo, N-0315 Oslo, Norway
}
\affiliation{
$^{3}$Department of Chemistry and Center for Materials Science and Nanotechnology, University of Oslo, Oslo 0371, Norway
}

\date{\today}
\begin{abstract}

The melting curve of pure silica (SiO$_2$) was determined using {\it ab initio} density functional theory together with the solid-liquid coexisting approach, thermodynamic integration and the Z method. The melting curves are consistent with a smooth slow increase in a large region from 50 GPa (dT/dP $\approx$ 15 K/GPa) to about 500 GPa (dT/dP $\approx$ 5 K/GPa) without any abrupt changes at around 120 GPa and 300 GPa as seen in some recent experimental and computational studies. The topography of the melting curve above 50 GPa is consistent with a gradual change in the distribution of the Si coordination numbers in the liquid state and the absence of large changes in the density following solid-solid phase transitions. The pair distribution functions show that the structural correlation in the liquid is mainly short-ranged and that the Si-O bond is stiff. The densification of the melt structure with pressure above 50 GPa is therefore due to an increase in 7- and 8-fold coordinated silicon.

\end{abstract}
\maketitle
\end{CJK*} 
\section{\label{sec:level1}Introduction}

Silica is a reference inorganic material with a wide range of applications in the chemical industry and one of the main components in terrestrial planets' crust and interior. In spite of its chemical simplicity, the SiO$_2$ phase diagram is surprisingly rich and a number of experimental as well as computational studies have attempted to draw subsolidus phase-relations and melting curves to very high pressures consistent with those in the Earth's deep interior~\cite{Usui2010, Das2020, Akins2002, Grocholski2013, Murakami2003, Kuwayama2011, Oganov2005} and beyond~\cite{Millot2015, Gonzales2016a, Gonzalez2016b}. These phase-relations are not only important for understanding the solid Earth's evolution from a very early stage following the magma ocean crystallization, but constraining the SiO$_2$ melting curve to very high- or even ultra-high pressure may provide insight into the formation of other rocky planets such as many super-Earths. Indeed since MgSiO$_3$ melt decomposes in  SiO$_2$ and MgO at pressures above 300 GPa~\cite{Boates2013}, liquid SiO$_2$ may be the most abundant phase in some of these large rocky planets' deep interiors. High pressure silica melting also plays an important role in the Earth's core dynamics since it has been suggested that silica may have crystallized from a Si-saturated proto-core during a chemical exchange with a basal magma ocean~\cite{Hirose2017, Tronnes2019}. 

Although silica phase-relations have undergone several revisions experimentally and computationally, its melting curve remains poorly constrained; in particular at very high pressure consistent with those in the Earth's core and beyond where experimental reports are scarce. A recent shock experiment study by Millot {\it et al}~\cite{Millot2015} reported a melting curve of $T_m(P)=1968.5+307.8\times P^{0.485}$ up to $\sim$ 500 GPa, suggesting that silica melts at around 5000  and 7000 K at pressures of the core-mantle boundary and the Earth's inner core respectively.

More recently, a Diamond Anvil Cell (DAC) study by Andrault {\it et al} \cite{Andrault2022, Andrault2020} reported melting points up to about 150 GPa. Their melting temperatures were markedly higher  ($\approx$ 6000 K at CMB pressure)than those of Millot {\it et al}~\cite{Millot2015}  and a very steep and abrupt change in the Clapeyron slope at $\sim$ 120 GPa was inconsistent with that seen of Ref.\cite{Millot2015}. Two solid-liquid coexisting (two-phase) molecular dynamics simulations using classical interatomic potentials~\cite{Belonoshko1995} and DFT \cite{Usui2010}  both report even higher melting temperatures (up to around 160 GPa) than those from the DAC experiment study, but none of these studies found any rapid change in the Clapeyron slope around 120 GPa. Moreover, a recent computational study~\cite{Gonzales2016a, Gonzalez2016b} using the Z method together with density functional theory reported a melting curve to ultra high pressure ($\sim$ 6000 GPa) consistent with those in the core of the gas-giants and massive super-Earths (where SiO$_2$ may be the main component).  Although their melting curve is in line with previous computational work up to around 150 GPa, the curve flattens markedly at around 200 GPa and is nearly flat until 300 GPa where it abruptly climbs steeply to almost 9000 K at 400 GPa. By contrast, the experimental study by Millot {\it et al}~\cite{Millot2015} did not capture any such anomalies. Clearly, there are many outstanding inconsistencies in the reported silica melting temperatures at high pressure.

Changes in the topography of the SiO$_2$ melting curve is tightly linked to any abrupt changes in the solid density and liquid structure with pressure. For example, the rapid increase in the melting curve of SiO$_2$  at a pressure of $\sim$ 14 GPa is attributable to a first order phase transition from coesite to stishovite and involves a change in the Si coordination environment from SiO$_4$ tetrahedra to SiO$_6$ octahedra accompanied with a huge increase in density of nearly~30\%~\cite{Akaogi2011}. Phases-transitions to higher-pressure polymorphs such as $\beta$-stishovite (CaCl$_2$-type), seifertite ($\alpha$-PbO$_2$-type) and pyrite are all accompanied by much smaller density changes of only a few percent~\cite{Das2020, Kuwayama2005, SolidCN2021, Li2023}. These phase transitions are therefore not expected to impact the topography of the melting curve to a similar extent as the coesite to stishovite transition. Likewise, the distribution of coordination numbers in the liquid is intimately connected to changes in the liquid density and entropy with pressure. As the distribution of different coordination polyhedra changes along with pressure, the liquid density and entropy will change too. Understanding melting processes thus requires an atomistic insight into the local structure of both solid and liquid phases at equilibrium.

Here we report results from a study of the melting curve of silica  up to 500 GPa using {\it ab initio} density functional theory together with three complementary methods to melting: The solid-liquid coexistence method, thermodynamic integration (TI) and the Z method (together with a waiting time analysis). The influence of local structure on melting is investigated to better understand the topography of the SiO$_2$ melting curve.

\section{\label{sec:level1}Method}
 
\subsection{\label{sec:level2}Two-phase coexistence method}

The solid-liquid (two-phase) coexisting approach aims to target the melting temperature by equilibrating a system where both a liquid and a solid are in mechanical contact in the simulation box. We launch {\it ab initio} Born-Oppenheimer MD (BOMD) runs from an initial configuration of atoms consisting of a solid and a liquid part with the temperature and volume kept fixed. As the system equilibrates, the fraction of the phases change along with the total pressure. If the initial pressure is higher than the equilibrium melting pressure  - with $\rho_{\text{solid}}  >  \rho_{\text{liquid}}$ where   $\rho_{\text{solid}}$ and $\rho_{\text{liquid}}$ are average solid and liquid densities respectively - the solid-liquid interface adjusts to increase the fraction of liquid during equilibration, thereby increasing the pressure in the direction of the equilibrium melting pressure. If the MD run is launched with a pressure that is lower than the melting pressure, the solid fraction of the box increases and the pressure decreases toward the melting curve during equilibration. As the system propagates, the pressure fluctuates around a fixed value, the melting pressure, and the boundaries separating the solid and liquid phases do not drift.  Under certain conditions, the system may melt or freeze quickly in a single simulation even with fairly large simulation boxes. However, we can still bracket the melting curve from a series of calculations at a given (T, P) and estimate the melting temperature from a  statistical distribution of the number of calculations that end up as a liquid or solid~\cite{Hong2016}. When, for example, half of the calculations at a given (T, P) end up in a liquid state, the simulation temperature is the melting temperature~\cite{Hong2016, Hernandez2022}. Here we use a slightly different approach where a sequence of MD runs are carried out at a given temperature with different volumes chosen to lie close to the expected liquid band. The melting pressure is then estimated from the midpoint between the MD run with the lowest pressure that froze and the MD run with the highest pressure that melted.  

\subsection{\label{sec:level2}Thermodynamic integration}
 
From the definition of melting equilibrium, the free energies of liquid and solid are equal at the melting point. We use thermodynamic integration to calculate  free energy of the liquid and solid state from an energy relationship between a reference state and an objective system 
\begin{equation}
    F_{\text{obj}}-F_{\text{ref}}=\int_{0}^{1} 	\left \langle U_{\text{obj}}(R)-U_{\text{ref}}(R)	\right \rangle_\lambda d\lambda.
\end{equation}
A carefully chosen sample of  $\langle U(\lambda)\rangle_\lambda$ calculated using {\it ab initio} BOMD simulations enables us to accurate calculation of the free energy of the objective system with sufficient accuracy.

An ideal gas is used as the reference state in order to calculate the free energy of liquid silica:
 \begin{equation}
    F_{\text{ref}} = F_{\text{ideal\ gas}}=-k_{\text{B}}T\text{ln}{\dfrac{V^N}{\Lambda^{3N}N!}}
\end{equation} where $\Lambda$ is the thermal de Broglie wavelength: \begin{equation}
    \Lambda=\dfrac{h}{\sqrt{2\pi m k_{\text{B}} T}}
\end{equation}
As discussed in previous studies~ (see e.g. Refs.\cite{TI2018, Melting2018, TI2019}),  accurately calculating $\langle U(\lambda)\rangle_\lambda$ requires a large number of $\lambda$ points in particular in the region where $\lambda \rightarrow 0$ (where the ensemble is ideal gas like). Instead of integration over $ \lambda \in [0, 1]$, the integration is performed over $ x \in [-1, 1]$ using 

\begin{equation}
    \lambda (x) = (\dfrac{x+1}{2})^{\dfrac{1}{1-k}}
\end{equation} where we choose $k$=0.8.

The $x(\lambda$) points are chosen using a Gauss-Lobatto quadrature which is shown to give a good balance between accuracy and computational cost.

For solids, our reference state is the free energy calculated within the quasi-harmonic approximation (QHA). That is
\begin{equation}
    F_{\text{ref}} = E_{0K}+F_{\text{har}} 
\end{equation}
where $E_{0K}$ is the energy of the relaxed crystal plus contributions from Zero point vibrations. The quasi-harmonic vibrational term can be expressed using the phonon density of states as follows
\begin{equation}
    F_{\text{har}}(T) = \frac{1}{2}\sum_{qj}{h\omega_{qj}}+k_{\text{B}}T\sum_{qj}\text{ln}[1-\text{exp}(-h \omega_{qj})/k_{\text{B}}T]
\end{equation}
where $q$ is the wave vector, $j$ is the band index and $\omega_{qj}$ is the phonon frequency of the phonon mode labelled by a set \{q,j\}. The phonon calculations are conducted using the finite displacement method as implemented in the Phonopy code\cite{Phononpy2015}. Thermodynamic integration is then used to calculate the anharmonic contribution to the solid from Eq. 1 using the Gauss-Legendre quadrature to integrate over $ x \in [-1, 1]$ by the changes of variables from $ \lambda \in [0, 1]$ using 
\begin{equation}
    \lambda (x) = \dfrac{x+1}{2}.
\end{equation}

Since  the reference state is the "static state" plus QHA corrections for the solid (Eq. 5), the difference between the reference and object functions reflects mainly the anharmonic contribution to the free energy.

\subsection{\label{sec:level2}Z method}

The Z method calculates the melting temperature from the relationship between a material's homogeneous melting temperature (superheating limit) and its equilibrium melting temperature~\cite{Belonoshko1995, Belonoshko2006, Alfe2011}. A sequence of molecular dynamics simulations are launched in the NVE ensemble~\cite{PBE96} at different initial temperatures ($T_{\text{ini}}$) to target the lowest total energy, $E_\text{h}$ (and highest temperature, $T_\text{h}$) where the solid melts. The maximum energy along the solid branch of the isochore is the same as the lowest energy along the liquid branch
\begin{equation}
E^{\text{sol}}(V, T_{\text{h}}) = E^{\text{liq}}(V,T_{\text{m}}).\label{waitingtime} 
\end{equation}
When the system melts at $T_{\text{h}}$, the temperature decreases to a distinct value, $T_\text{m}$, the equilibrium melting temperature, while the latent heat of melting is gradually converted to potential energy. The relationship between $T_\text{h}$ and $T_\text{m}$ is given by~\cite{Belonoshko2006, Braithwaite2019} 
\begin{equation}
   \frac{T_{\text{h}}}{T_{\text{m}}} - 1 = \frac{\Delta S_{\text m}}{C_V}\label{eq1}.
\end{equation}
Here $C_V$ is the heat capacity of the solid and $\Delta S_m$ is the melting entropy.

Since the waiting time for the solid to melt diverges when $T$ tends to $T_\text{h}$, the calculated melting temperature will always, in theory, represent an upper bound to the "true" melting temperature. To avoid extremely long MD runs in the vicinity of $T_\text{h}$, the melting temperature is instead often calculated from an extrapolation of the distributions of waiting times from a sequence of runs with $E = (E_\text{h} + \Delta E)$ using
\begin{equation}
  \langle \tau \rangle^{-1/2} = A(T_{\text{liq}} - T_{\text{m}})
  \label{Eqwait}
\end{equation}
where "A" is a parameter, $\tau$ is the waiting time for a given total energy and $T_{\text{liq}}$ is the temperature of the system after melting~\cite{Alfe2011}. When $\Delta E \to 0$, then $T_{\text{liq}} \to T_\text{m}$ and the melting temperature is found at the point of intersection where $\langle \tau \rangle^{-1/2} = 0$.  

Implementation of the Z method together with DFT typically uses {\it ab initio} BOMD in the NVE ensemble where the Fermi-Dirac electronic temperature is kept fixed along the entire trajectory. Although this ensures conserved dynamics using the Hellmann-Feynman forces to propagate the ions~\cite{Mermin1965, Wentzcovitch1992}, large changes in the temperature following melting and equilibration may introduce systematic errors in the calculated melting temperature. The difference in melting temperature when $T_\text{el} \approx T_\text{h}$ and $T_\text{el} \approx T_\text{m}$ - representing reasonable lower and  upper bounds respectively to $T_{\text{m}}$ - is about 200 K for SiO$_2$ at around 160 GPa and 6000 K~\cite{ZMethod2023}. Here we choose an electronic entropy close to the predicted melting temperature. Although this ensures a correct ion-electron interaction in the liquid state of the MD simulation, the electronic temperature is slightly too low before melting and the melting temperature will probably be overestimated by $\sim$ 100 K~\cite{ZMethod2023}.

\section{\label{sec:level1}Computational Details}

All calculations carried out in this work were performed using the Vienna {\it ab-initio} simulation package \cite{VASP93, VASP94} together with the GGA functional (parametrized using the Perdew-Burke-Ernzerhof (PBE) scheme)\cite{PBE96} to calculate the exchange-correlation contribution to the total energy implemented with projector augmented-waves (PAW)~\cite{PAW94, PAW99}. The valence electron configurations were 2$s^2p^4$ for O and $3s^23p^2$ for Si with core-radii of 1.95~\AA and 1.55~\AA respectively for the projector operator. Two-phase test calculations of thermodynamic (average pressure) and structural properties (PDFs) carried out using "hard" oxygen potential (with a core-radii of 1.228 \AA) were in excellent agreement with those calculated using the default potentials. For example, the average pressure for pyrite at 8000 K with "hard" O-potentials was within the error bars of that reported using default potentials and the PDFs were nearly indistinguishable.

A cut-off energy of 400 eV for the plane waves were used in the two-phase simulations. Test calculations carried out using a cut-off energy of 500 eV gave ensemble average properties (i.e. average pressures and PDFs) within the error bars reported using a 400 eV energy cut-off. Since a markedly smaller cell size was used in the Z method calculations compared to the two-phase calculations, the energy cut-off was slightly higher (700 eV). In the TI calculations, we analysed the convergence of the static energy from a sequence of runs from 500 eV to 1000 eV. We found that the total energy was sufficiently converged ($<$ 0.001 eV/atom) using an energy cut-off of 800 eV. This value is used for all the calculations carried out in the static limit as well as in the QHA runs. $4\times4\times4$ supercells are used in the evaluation of the harmonic vibrational contribution to the total free energy. In the MD simulations where the anharmonic contribution to the solid free energy was calculated, we found that an energy cut-off of 500 eV was sufficient. That is we found that the DFT energies calculated from MD runs with a higher energy cut-off of 800 eV differ by only about 0.00103 eV/atom compared to that calculated using an energy cut-off of 500 eV. The $\Gamma$-point only were used in all the MD runs used in the two-phase, TI and Z Method calculations,  whereas a $4\times4\times4$ Monkhorst-Pack mesh was used in the TI calculations for the solid carried out in the static limit.

In all NVT MD runs the electronic entropy was included using a Fermi-Dirac smearing scheme with a width of $k_\text{B}T$ \cite{Mermin1965}. As discussed above, special attention was given to the choice of electronic temperature in the Z method calculations because of the large temperature drop accompanied melting in the simulations.  We evaluated the melting temperatures using three different electronic temperatures ranging from 5500 K to 7500 K for seifertite and 6000 K to 8000 K for pyrite. The electronic temperature which is closest to  $T_{\text{m}}$ is chosen in the waiting time analysis and values used are reported in the Supplemental Information (SI).

The molecular dynamics runs used in the two-phase calculations and the TI free energy calculations were carried out using a Nos\'e-Hoover thermostat with a time step of 1 (TI) or 2 fs. A smaller time-step of 0.5 fs was used in the Z method calculations. 

For the liquid TI simulations, $\langle U(\lambda)\rangle_\lambda$  was evaluated with an integration of 8 $\lambda $ points using a Gauss-Lobatto quadrature. The energy difference between 8 $\lambda $ points and 10 $\lambda $ points evaluated in a 162 atoms simulation box is only $\sim 0.0008$ eV/atom suggesting that 8  $\lambda $ points is sufficient (see the SI Table S2 and Fig. S1 where we plot the convergence of the free energy with the number of $\lambda$ points). 

The two-phase coexistence calculations were carried out using a 486 atoms simulation cell for stishovite and $\beta$-stishovite, whereas a 648 atoms cell was used for seifertite and pyrite. These simulation boxes are made from units constructed from the primitive SiO$_2$ unit cell of a given solid phase. The liquid portion of the two-phase simulation box was constructed by first heating to above the homogeneous melting temperature followed by a decrease in temperature to $T_{\text{ini}}$. The solid portion was then heated until $T_{\text{ini}}$ and two equal volumes of the solid and liquid sub-unites are glued together~\cite{Hernandez2022}. We then launch the two-phase MD simulations at $T_{\text{ini}}$ following the procedure discussed in the Theory section. All two-phase calculations ran for at least 5 ps and sometimes for more than 40 ps before they froze or melted. After fully melted (or crystallized) the MD runs continued for at least 1 ps to calculate the equilibrium pressure with sufficient accuracy. Equilibrium pressures and relaxation times for melting or freezing are reported in Table S1. 

We also investigated the sensitivity to melting using different surface cut-off schemes to represent the solid-liquid phase boundary. Comparison from calculations carried out on stishovite using a (001) vs a (100) crystallographic cut-off plane was found to give very similar melting temperatures.

In total, we carried out more than 100 two-phase calculations in the pressure range from 50 to 500 GPa and details from most of these (including those used in the calculation of the melting curve) are reported in Table S1 in SI. We used a large pressure overlap ($\sim$ 25 GPa) for the different solid phases to ensure that the correct (equilibrium) solid phases where used to calculate the melting curve. The expected stability range of the solids are estimated from an extrapolation of the Clapeyron-slopes determined from previous recent computational and experimental studies (see e.g.~\cite{Das2020} and references therein).

\section{\label{sec:level1}Results and Discussion}


\begin{figure*}
\includegraphics[width=0.94\textwidth]{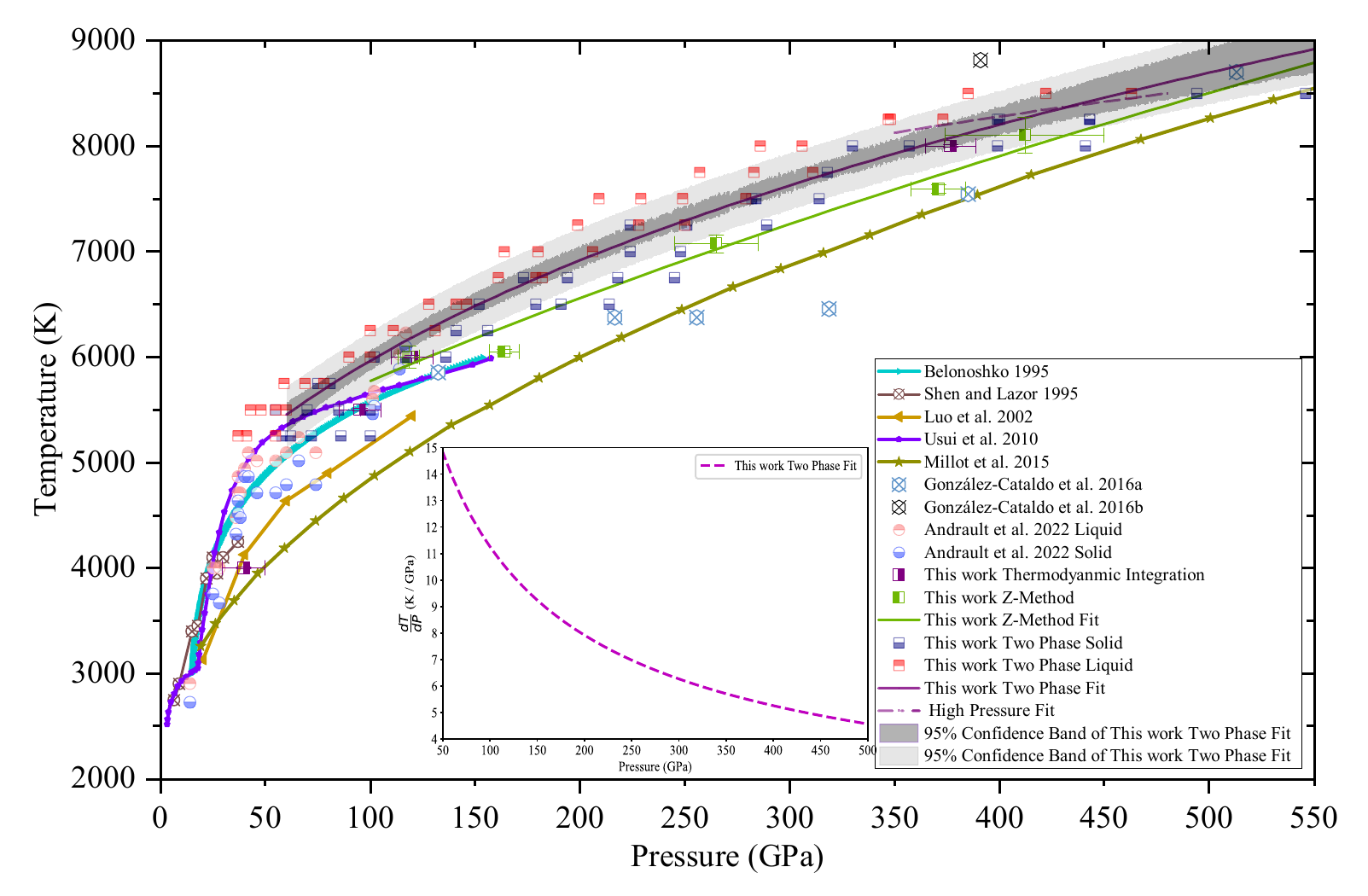}
\caption{Melting points and melting curves of SiO$_2$ from this work and the literature. We fitted Simon-Glatzel melting equations using $T_{\text{melting}}(P)=T_0 \times (1+ \dfrac{P-P_0}{a})^b$ to the melting points from the Z method and two-phase calculations, which gave the following parameters (two-phase calculations): $T_0=6.018, P_0=-202.338, a=4.736 \times 10^{-4}, b=0.5126 $; and (Z method calculations): $T_0=6.462, P_0=-256.011, a=6.474\times10^{-4}, b=0.5141$. For the two-phase calculations the data points used in the fitting were the mean values of the lowest liquid and highest solid phases at a given temperature. All data points from the two-phase calculations are also reported in Table S1 in SI . The inset shows the steepness of the melting curves from the two-phase coexistence calculations.}
\label{meltingcurves}
\end{figure*}

In Fig.~\ref{meltingcurves} we collect calculated melting temperatures using the Z method, TI and two-phase methods together with previously reported results in the literature. We focus here on the functional form of the SiO$_2$ liquidus above about 50 GPa as the steep increase in the melting curve between  14 and 50 GPa, due to the coesite-stishovite transitions, is well documented \cite{CoeTrans1969, CoeTrans1976, CoeTrans1995, CoeTrans1996, CoeTrans2015, CoeTrans2022}. Above about 50 GPa our melting curves flattens significantly and the discrepancy between the different melting temperatures and curves reported in the literature span more than 1000 K in a large pressure interval. The Z method and two-phase melting curves are in overall very good agreement with the melting points calculated using TI at around 100, 120 and 377 GPa, but the calculated TI melting temperature at around 50 GPa is somewhat lower compared to the melting curve from our two-phase-  and Z method calculations as well as previous computational work~\cite{Usui2010, Gonzales2016a}. This discrepancy may be due to the possibly large error bars in the solid and liquid free energy curves using TI, and the extremely steep melting curve below 50 GPa.  

The melting curve from the two-phase calculations have a dT/dP $\approx$ 14 K/GPa at around 50 GPa and the dT/dP is about 5 K/GPa at 500 GPa without any abrupt changes in the functional form. A possible flattening anomaly above about 300 GPa has been indicated in the figure as a dashed line. Such a small anomaly could possibly be due to an increase in the liquid electronic entropy rather than the configurational liquid entropy since there are no changes in the liquid structure (i.e. distribution of coordination numbers) that points to a configurational stabilization of the liquid state above 200 GPa. That is in a large pressure interval from 200 to 400 GPa, the average CN for Si increases by less than 0.2 from 5.7 to 5.9. 

Although the Z method and two-phase melting curves are in overall good agreement in the entire pressure range studied, the Z method curve is slightly lower by about 300 K and the Simon-Glatzel fit to the Z method data indicates a slightly flatter curve at lower pressures compared the two-phase melting curve.  Since the same parameters for the electronic wave function were used in the two methods a possible explanation for the small discrepancy could be due to statistics, cell size effects or choice of electronic entropy in the NVE MD runs (Z method). 

In particular, the choice of electronic entropy is not evident in MD runs carried out in the NVE ensemble because the Mermin free energy plus the ionic kinetic is a conserved quantity with the forces being propagated using Hellmann-Feynman dynamics~\cite{Wentzcovitch1992}. This means that the electronic temperature,  $T_{\text{el}}$, is {\it kept fixed} along the entire trajectory. A fixed electronic temperature, however,  may introduce errors in the calculated melting temperature because it is unable to correctly capture electron-ion dynamics in both the solid and liquid portion of the MD simulation since the temperature difference between these states is large. If an MD run is launched in the NVE ensemble with $T_{\text{el}}$ chosen near the liquid temperature, then $T_{\text{el}}$ will be much lower than the ensemble temperature in the solid state (before melting). A too low electronic entropy, however, may favour the stabilisation of the solid and prevent melting. This will affect the estimated homogeneous melting temperature and also the "waiting time" for a solid to melt. Thus the calculated melting temperature may be too high. If an MD run is launched with a much higher electronic temperature - for example, chosen to lie close to the temperature of the solid state  - a physically reasonable electronic-ionic interaction is only ensured before melting. This is because once the solid melts at a constant volume and total energy, the temperature drops by (1 - $T_{\text{m}}/T_{\text{h}})\times T_{\text{m}} \approx$ 2000 K (see Fig. S6) and the electronic entropy will be much {\it higher} than the liquid temperature \cite{ZMethod2023}. This generally favours an entropic stabilisation of the liquid and the melting temperature may be too low. Here we choose  $T_{\text{el}} \approx T_{\text{m}}$ in the Z method calculations, indicating that the melting temperature may be slightly overestimated by about 100 K for SiO$_2$ at T$\sim$ 6000 K~\cite{ZMethod2023} (see also Fig S6 in SI). Other possible sources to the small discrepancy between these two curves could be due to the formation of defects in the solid state accompanying freezing. Such defects were not seen in the Z method runs when the box melted homogeneously, but we noted that such defects sometimes form in the two-phase calculations (as marked in Table S1 in SI). Anyhow, the overall good agreement between all 3 methods employed in this work is encouraging and suggests the melting curve is well represented by the smooth Simon-Glatzel fits shown in Fig. 1.

The good agreement with our melting curves and that reported by Usui and Tschuchia ~\cite{Usui2010} (plotted in Fig. 1) is therefore possibly fortuitous because a very small two-phase simulation box containing only 96 atoms was used in Ref.~\cite{Usui2010}. Such a box has a boundary of similar size (or larger) than the solid and liquid portions of the simulation box and extensive statistics is needed to precisely determine the melting temperature using such a small two-phase cell~\cite{Hong2016, Hernandez2022}. Moreover, the slope in Ref.~\cite{Usui2010}  above about 60 GPa is also markedly less steep than our melting curves suggesting that it will lie markedly lower than that of Millot {\it et al}~\cite{Millot2015} if extrapolated above 200 GPa.

Also worth noting is that our results are inconsistent with the Z method calculations reported by Gonzales {\it et al}~\cite{Gonzales2016a, Gonzalez2016b}. Whereas our curves show a smooth increase in the pressure region between 200 and 400 GPa, the melting points by Gonzales {\it et al}~\cite{Gonzales2016a, Gonzalez2016b} indicate a nearly flat curvature from 200 GPa to 350 GPa followed by an abrupt increase in temperature by more than 2500 K at around 400 GPa. This anomaly was explained in Ref.~\cite{Gonzales2016a} by the seifertite to pyrite transition which has been suggested to take place at around 280 GPa according to an experimental high-pressure study~\cite{Kuwayama2005}. Computational studies, however, consistently show that this transition takes place at a much lower pressure i.e. at around 200 GPa or slightly less~\cite{Oganov2005, Das2020}. For example, highly accurate hybrid functionals to DFT together with lattice dynamics predicts that the transition to pyrite takes place at  around 200 GPa~\cite{Das2020} indicating that the experimental work by Kuwayama {\it al}.~\cite{Kuwayama2005} may be poorly constrained. Indeed the outcome from four different experiments reported in Ref.~\cite{Kuwayama2005} all indicate a nearly vertical, but weakly constrained, dP/dT slope, and the low experimental temperatures ($<$ 2000 K) could therefore explain the delay in the transition pressure due to slow kinetics. Such a "delayed" phase transition has also been observed for other phase transitions in silica, for example, the $\beta$-stishovite to seifertite transition (see e.g. Ref.~\cite{Murakami2003}), resulting in a too high transition-pressure before the high-pressure phase was visible in the X-Ray diffraction patterns. 

Moreover, the volume drop accompanying the \st{$\beta$-stishovite to} seifertite to pyrite transition is expected to be around 4 \%~\cite{Kuwayama2005, Das2020} which is probably too small to explain the rapid increase in the melting curve close to 400 GPa as suggested in Refs.~\cite{Gonzales2016a, Gonzalez2016b}. An explanation for such a steep slope can possibly be associated with the choice of electronic temperature in the DFT calculations and/or slow equilibration in the MD NVE runs close to the homogeneous melting temperature which may prevent melting. To investigate the origin of the discrepancy between our melting curves and the melting point reported in Ref.~\cite{Gonzalez2016b} at 8826 K and 391 GPa using the Z method, we attempted to reproduce their high temperature melting anomaly. We thus carried out MD simulations with an initial temperature of 25000 K and   $T_{\text{el}}$ = 7500 K with all atoms initially located at their equilibrium positions. This initial temperature is slightly lower than that used in Ref.~\cite{Gonzalez2016b}) which is 26000 K. With these parameters, pyrite melts quickly ($<$ 5 ps in all our 11 parallel MD runs) reaching an average liquid temperature and pressures of 8608 K and 383 GPa. This is quite close to the liquid point estimated in Ref.~\cite{Gonzalez2016b}  (i.e. 391 GPa and 8826 K).  Although no electronic temperatures were reported in Ref~\cite{Gonzalez2016b}, reasonable choices in the range 6000-9000 K is expected to influence the melting temperature to less than 300 K~\cite{ZMethod2023}. Rapid melting in our MD simulations with  $T_{\text{ini}}$  = 25000 K, however, indicate that the liquid temperature with this choice of initial conditions is far higher than the equilibrium melting temperatures. When we decrease the initial temperature to as low as 22000 K the system still melts, but the waiting time before melting was more than 120 ps in all our 11 parallel MD runs, and the liquid temperature was 7747 K at 372 GPa. This is only about 100 K higher than the equilibrium melting temperature extracted from a waiting time analysis. This confirms that liquid temperature in  MD runs with $T_{\text{ini}}$ = 26000 K~\cite{Gonzalez2016b} - and assuming that the atoms are initially distributed over their ideal positions with a reasonable choice of T$_{\text{el}}$ - is $\approx$ 1000 K higher than the equilibrium melting temperature at 390 GPa. 

Our MD simulations may also provide insight into the kinetics of shock-compressed silica and shed some light on the discrepancy between our SiO$_2$ equilibrium melting curve and that extracted from shock experiment by Millot {\it et al}~\cite{Millot2015}. Shock Hugoniots appear to lie at different branches corresponding to amorphous or crystalline states depending on experimental methodologies and time scales~\cite{Shen2016}. If the liquid is unable to relax to the crystalline state at the time scale of the shock pressure, one might expect that the Hugoniot may lie at lower temperatures compared to that of a fully relaxed (crystalline) state. Indeed, large scale atomistic MD simulations carried out by Shen {\it et al}~\cite{Shen2016} show that points on the amorphous Hugoniot will shift to the crystallized branch at longer timescales. The discrepancy between our equilibrium melting curve and that predicted from a chock study by Millot {\it et al}~\cite{Millot2015} could therefore be due to the incomplete crystallization of stishovite or amorphous shock compressed silica.

The total and Si-O partial pair distribution functions (PDFs) plotted in Fig.~\ref{PDF} show that the correlation length in the liquid state is mainly short-ranged ($< 5$~\AA) pointing to a high liquid entropy quantified by the relative distribution of Si CNs. Comparison of the Si-O partial PDFs at different pressures along the liquidus show surprisingly little changes in the Si-O bond length: The first peak in the Si-O partial PDF is located at 1.65~\AA~at 6000 K and 100 GPa whereas Si-O bonds is only about 0.1~\AA~shorter at  8250 K and 373 GPa. The driving force to increasing liquid density (and entropy) is therefore due to changes in local coordination and/or electronic entropy. In particular, the electronic entropy could be a main source to the very flat melting curve above 300 GPa since there is very little changes in the liquid structure in this region reflected by a very small increase in the average Si CN from 5.8 (300 GPa) to 5.9 (400 GPa) with pressure.

The partial PDFs of the solids are in overall good agreement with previous studies (see e.g. Ref.~\cite{Kuwayama2011}). For example the comparison of the nearest neighbouring O-O distances is at about 2.2~\AA~at 214 GPa in good agreement with that seen in Ref at slightly lower pressure.~\cite{Kuwayama2011}. In Fig. S7 in SI, we collect XRD patterns of the liquid at two pressures (76 GPa and 275 GPa). The pattern is consistent with a previous study~\cite{liqsilica2018, HPLiquid} containing a broad peak that shifts to higher values of 2$\theta$ with increasing pressure in line with the increasing densification of liquid with short-ranged correlation.

\begin{figure*}
\includegraphics[width=0.94\textwidth]{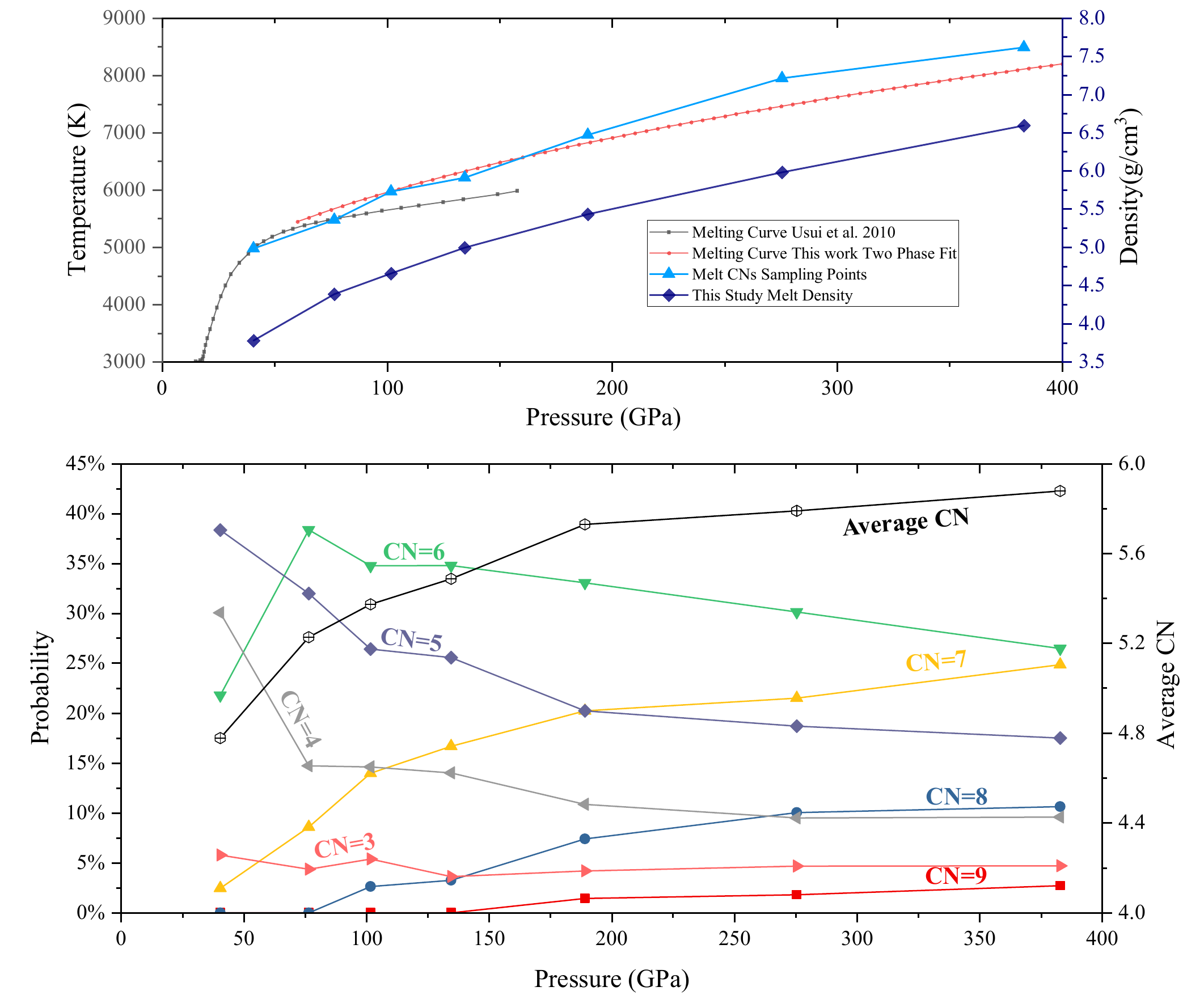}
\caption{Top and bottom graphs show the changes in the density and Si coordination numbers respectively in liquid SiO$_2$ near the melting curve. The coordination numbers are evaluated with a Near-Neighbor algorithm~\cite{CrystalNN2020} which uses a Voronoi decomposition scheme to determine the probability of various coordination numbers and then select the one with the highest probability.}
\label{CNL}
\end{figure*}

\begin{figure*}
\includegraphics[width=0.48\textwidth]{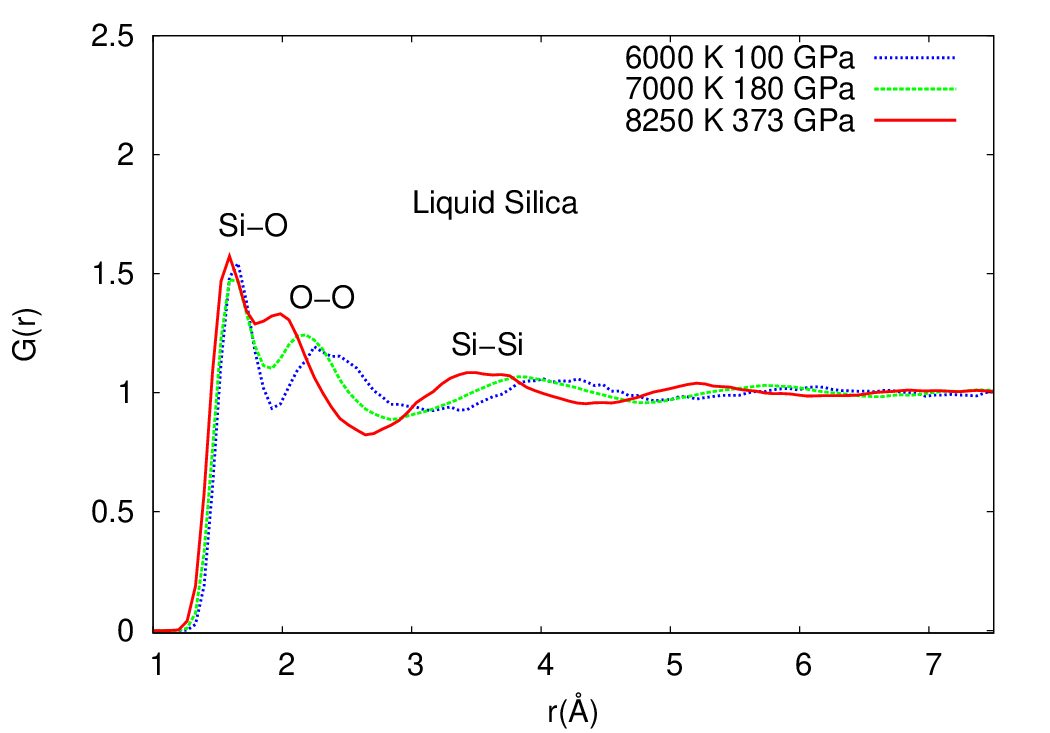}
\includegraphics[width=0.48\textwidth]{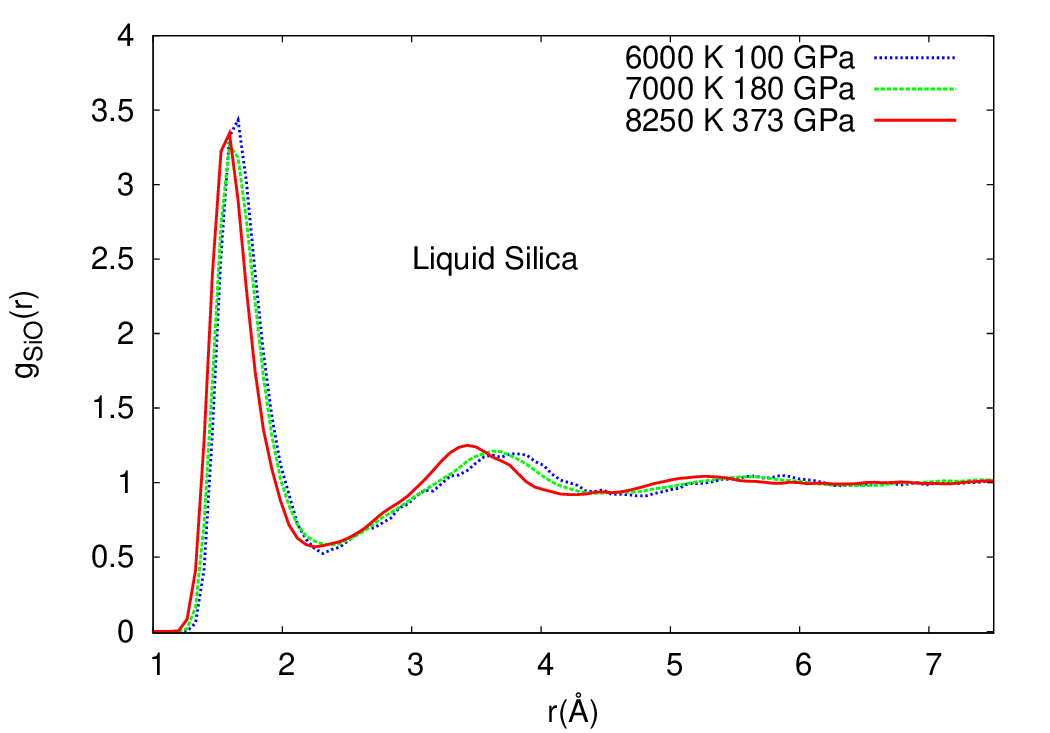}
\includegraphics[width=0.48\textwidth]{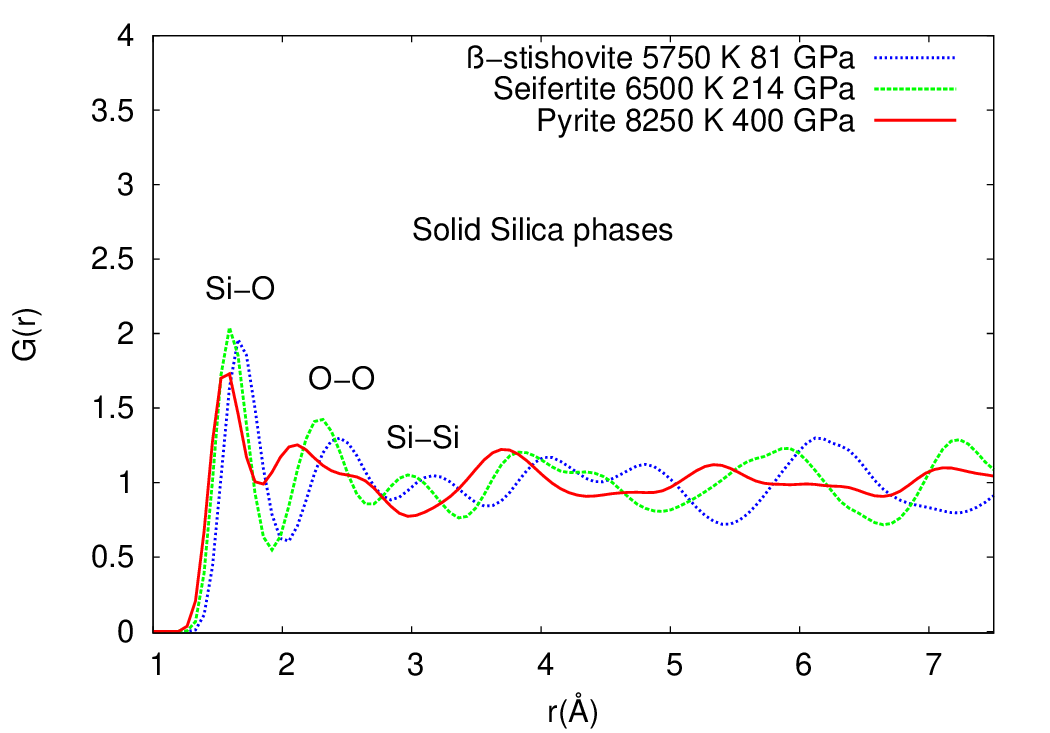}
\includegraphics[width=0.48\textwidth]{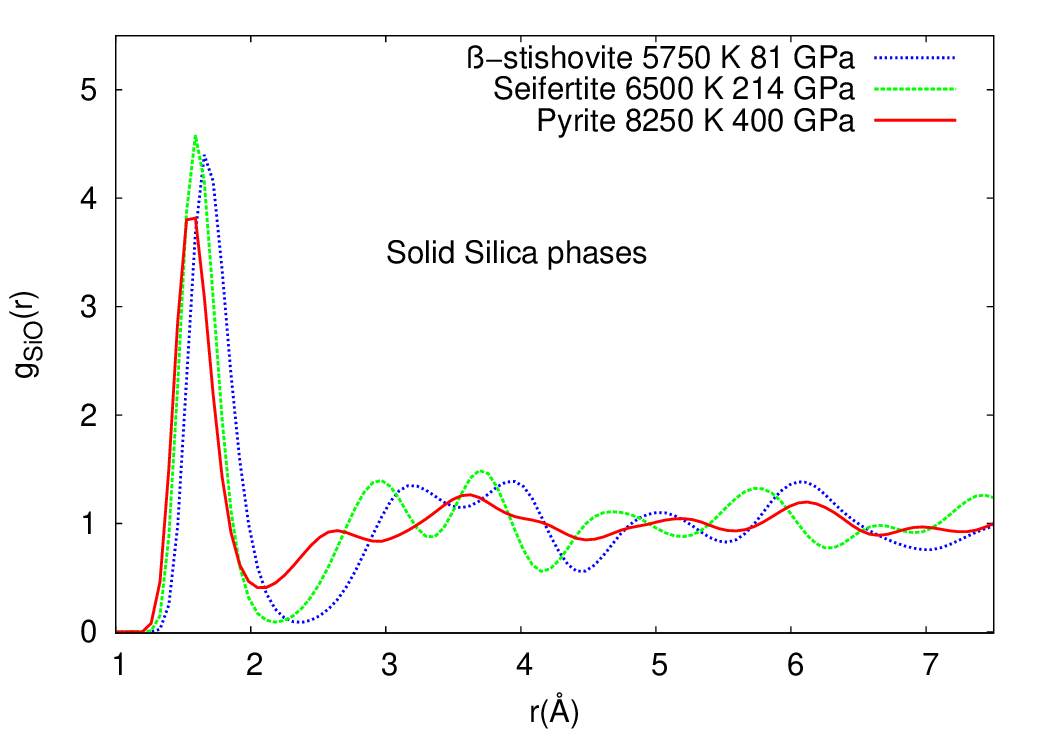}
\caption{Top left and right are the total and partial (Si-O) PDFs for liquid silica, whereas the bottom left and right are the corresponding (total and partial) PDFs for the solid phases.}
\label{PDF}
\end{figure*}

\section{Conclusions}

We have constrained the melting curve of pure silica from about 50 to 500 GPa using DFT at the level of GGA together with three complementary approaches to melting: the solid-liquid coexistence approach, thermodynamic integration and the Z method. The melting curves/points from the three different approaches are in overall very good agreement. After an abrupt increase following the transition to stishovite, the two-phase melting curves flatten markedly at about 50 GPa and increase smoothly from about 50 GPa with a dT/dP slope of $\approx$ 15 K/GPa to about 500 GPa with dT/dP $\approx$ 5 K/GPa. We do not see any evidence for an abrupt change at around 120 GPa which was seen in a recent experimental study by~\cite{Andrault2020, Andrault2022} nor the flattening observed from simulations using the Z method reported in Ref.~\cite{Gonzales2016a} at 200 GPa followed by an abrupt increase of almost 3000 K in the pressure window from 300 to 400 GPa. The topography of the melting curves from 50 to 500 GPa is consistent with a gradual change in the distribution of the Si coordination numbers in the liquid state and the absence of large changes in the density following solid-solid phase transitions. The pair distribution functions show that the correlations in the liquid structure is mainly short-ranged and that the Si-O bond decreases by less than 0.2 \AA ~ along the melting line from 100 to 400 GPa. The densification of the melt structure with pressure above 50 GPa is therefore mainly due to a gradual increase in 7-, 8- and 9-fold coordinated silicon and a gradual decrease in Si with 6 or fewer oxygens in the first coordination shell.

\begin{acknowledgments}
We thank Jean-Alexis Hernandez for useful input to the manuscript. We acknowledge financial support from the Research Council of Norway through its Centres of Excellence scheme, project number 223272 (CEED) and project number 332523 (PHAB). Computational resources were provided by the Norwegian infrastructure for high-performance computing (NOTUR, Sigma-2, Grants NN9329K and NN2916K). 
\end{acknowledgments}


\bibliography{SiO2}

\begin{thebibliography}{47}%
\makeatletter
\providecommand \@ifxundefined [1]{%
 \@ifx{#1\undefined}
}%
\providecommand \@ifnum [1]{%
 \ifnum #1\expandafter \@firstoftwo
 \else \expandafter \@secondoftwo
 \fi
}%
\providecommand \@ifx [1]{%
 \ifx #1\expandafter \@firstoftwo
 \else \expandafter \@secondoftwo
 \fi
}%
\providecommand \natexlab [1]{#1}%
\providecommand \enquote  [1]{``#1''}%
\providecommand \bibnamefont  [1]{#1}%
\providecommand \bibfnamefont [1]{#1}%
\providecommand \citenamefont [1]{#1}%
\providecommand \href@noop [0]{\@secondoftwo}%
\providecommand \href [0]{\begingroup \@sanitize@url \@href}%
\providecommand \@href[1]{\@@startlink{#1}\@@href}%
\providecommand \@@href[1]{\endgroup#1\@@endlink}%
\providecommand \@sanitize@url [0]{\catcode `\\12\catcode `\$12\catcode `\&12\catcode `\#12\catcode `\^12\catcode `\_12\catcode `\%12\relax}%
\providecommand \@@startlink[1]{}%
\providecommand \@@endlink[0]{}%
\providecommand \url  [0]{\begingroup\@sanitize@url \@url }%
\providecommand \@url [1]{\endgroup\@href {#1}{\urlprefix }}%
\providecommand \urlprefix  [0]{URL }%
\providecommand \Eprint [0]{\href }%
\providecommand \doibase [0]{https://doi.org/}%
\providecommand \selectlanguage [0]{\@gobble}%
\providecommand \bibinfo  [0]{\@secondoftwo}%
\providecommand \bibfield  [0]{\@secondoftwo}%
\providecommand \translation [1]{[#1]}%
\providecommand \BibitemOpen [0]{}%
\providecommand \bibitemStop [0]{}%
\providecommand \bibitemNoStop [0]{.\EOS\space}%
\providecommand \EOS [0]{\spacefactor3000\relax}%
\providecommand \BibitemShut  [1]{\csname bibitem#1\endcsname}%
\let\auto@bib@innerbib\@empty
\bibitem [{\citenamefont {Usui}\ and\ \citenamefont {Tsuchiya}(2010)}]{Usui2010}%
  \BibitemOpen
  \bibfield  {author} {\bibinfo {author} {\bibfnamefont {Y.}~\bibnamefont {Usui}}\ and\ \bibinfo {author} {\bibfnamefont {T.}~\bibnamefont {Tsuchiya}},\ }\bibfield  {title} {\bibinfo {title} {Ab initio two-phase molecular dynamics on the melting curve of \text{SiO$_2$}},\ }\href {https://doi.org/10.1007/s12583-010-0126-9} {\bibfield  {journal} {\bibinfo  {journal} {Journal of Earth Science}\ }\textbf {\bibinfo {volume} {21}},\ \bibinfo {pages} {801} (\bibinfo {year} {2010})}\BibitemShut {NoStop}%
\bibitem [{\citenamefont {Das}\ \emph {et~al.}(2020)\citenamefont {Das}, \citenamefont {Mohn}, \citenamefont {Brodholt},\ and\ \citenamefont {Tr{\o}nnes}}]{Das2020}%
  \BibitemOpen
  \bibfield  {author} {\bibinfo {author} {\bibfnamefont {P.~K.}\ \bibnamefont {Das}}, \bibinfo {author} {\bibfnamefont {C.~E.}\ \bibnamefont {Mohn}}, \bibinfo {author} {\bibfnamefont {J.~P.}\ \bibnamefont {Brodholt}},\ and\ \bibinfo {author} {\bibfnamefont {R.~G.}\ \bibnamefont {Tr{\o}nnes}},\ }\bibfield  {title} {\bibinfo {title} {High-pressure silica phase transitions: Implications for deep mantle dynamics and silica crystallization in the protocore},\ }\href {https://doi.org/10.2138/am-2020-7299} {\bibfield  {journal} {\bibinfo  {journal} {American Mineralogist}\ }\textbf {\bibinfo {volume} {105}},\ \bibinfo {pages} {1014} (\bibinfo {year} {2020})}\BibitemShut {NoStop}%
\bibitem [{\citenamefont {Akins}\ and\ \citenamefont {Ahrens}(2002)}]{Akins2002}%
  \BibitemOpen
  \bibfield  {author} {\bibinfo {author} {\bibfnamefont {J.~A.}\ \bibnamefont {Akins}}\ and\ \bibinfo {author} {\bibfnamefont {T.~J.}\ \bibnamefont {Ahrens}},\ }\bibfield  {title} {\bibinfo {title} {Dynamic compression of \text{SiO$_2$}: A new interpretation},\ }\href {https://doi.org/10.1029/2002GL014806} {\bibfield  {journal} {\bibinfo  {journal} {Geophysical Research Letters}\ }\textbf {\bibinfo {volume} {29}},\ \bibinfo {pages} {31} (\bibinfo {year} {2002})}\BibitemShut {NoStop}%
\bibitem [{\citenamefont {Grocholski}\ \emph {et~al.}(2013)\citenamefont {Grocholski}, \citenamefont {Shim},\ and\ \citenamefont {Prakapenka}}]{Grocholski2013}%
  \BibitemOpen
  \bibfield  {author} {\bibinfo {author} {\bibfnamefont {B.}~\bibnamefont {Grocholski}}, \bibinfo {author} {\bibfnamefont {S.-H.}\ \bibnamefont {Shim}},\ and\ \bibinfo {author} {\bibfnamefont {V.~B.}\ \bibnamefont {Prakapenka}},\ }\bibfield  {title} {\bibinfo {title} {Stability, metastability, and elastic properties of a dense silica polymorph, seifertite},\ }\href {https://doi.org/10.1002/jgrb.50360} {\bibfield  {journal} {\bibinfo  {journal} {Journal of Geophysical Research: Solid Earth}\ }\textbf {\bibinfo {volume} {118}},\ \bibinfo {pages} {4745} (\bibinfo {year} {2013})}\BibitemShut {NoStop}%
\bibitem [{\citenamefont {Murakami}\ \emph {et~al.}(2003)\citenamefont {Murakami}, \citenamefont {Hirose}, \citenamefont {Ono},\ and\ \citenamefont {Ohishi}}]{Murakami2003}%
  \BibitemOpen
  \bibfield  {author} {\bibinfo {author} {\bibfnamefont {M.}~\bibnamefont {Murakami}}, \bibinfo {author} {\bibfnamefont {K.}~\bibnamefont {Hirose}}, \bibinfo {author} {\bibfnamefont {S.}~\bibnamefont {Ono}},\ and\ \bibinfo {author} {\bibfnamefont {Y.}~\bibnamefont {Ohishi}},\ }\bibfield  {title} {\bibinfo {title} {Stability of \text{CaCl$_2$-type} and \text{$\alpha$-PbO$_2$-type} \text{SiO$_2$} at high pressure and temperature determined by in-situ x-ray measurements},\ }\bibfield  {journal} {\bibinfo  {journal} {Geophysical Research Letters}\ }\textbf {\bibinfo {volume} {30}},\ \href {https://doi.org/10.1029/2002GL016722} {10.1029/2002GL016722} (\bibinfo {year} {2003})\BibitemShut {NoStop}%
\bibitem [{\citenamefont {Kuwayama}\ \emph {et~al.}(2011)\citenamefont {Kuwayama}, \citenamefont {Hirose}, \citenamefont {Sata},\ and\ \citenamefont {Ohishi}}]{Kuwayama2011}%
  \BibitemOpen
  \bibfield  {author} {\bibinfo {author} {\bibfnamefont {Y.}~\bibnamefont {Kuwayama}}, \bibinfo {author} {\bibfnamefont {K.}~\bibnamefont {Hirose}}, \bibinfo {author} {\bibfnamefont {N.}~\bibnamefont {Sata}},\ and\ \bibinfo {author} {\bibfnamefont {Y.}~\bibnamefont {Ohishi}},\ }\bibfield  {title} {\bibinfo {title} {Pressure-induced structural evolution of pyrite-type \text{SiO$_2$}},\ }\href {https://doi.org/10.1007/s00269-011-0431-6} {\bibfield  {journal} {\bibinfo  {journal} {Physics and Chemistry of Minerals}\ }\textbf {\bibinfo {volume} {38}},\ \bibinfo {pages} {591} (\bibinfo {year} {2011})}\BibitemShut {NoStop}%
\bibitem [{\citenamefont {Oganov}\ \emph {et~al.}(2005)\citenamefont {Oganov}, \citenamefont {Gillan},\ and\ \citenamefont {Price}}]{Oganov2005}%
  \BibitemOpen
  \bibfield  {author} {\bibinfo {author} {\bibfnamefont {A.~R.}\ \bibnamefont {Oganov}}, \bibinfo {author} {\bibfnamefont {M.~J.}\ \bibnamefont {Gillan}},\ and\ \bibinfo {author} {\bibfnamefont {G.~D.}\ \bibnamefont {Price}},\ }\bibfield  {title} {\bibinfo {title} {Structural stability of silica at high pressures and temperatures},\ }\href {https://doi.org/10.1103/PhysRevB.71.064104} {\bibfield  {journal} {\bibinfo  {journal} {Phys. Rev. B}\ }\textbf {\bibinfo {volume} {71}},\ \bibinfo {pages} {064104} (\bibinfo {year} {2005})}\BibitemShut {NoStop}%
\bibitem [{\citenamefont {Millot}\ \emph {et~al.}(2015)\citenamefont {Millot}, \citenamefont {Dubrovinskaia}, \citenamefont {\u{C}ernok}, \citenamefont {Blaha}, \citenamefont {Dubrovinsky}, \citenamefont {Braun}, \citenamefont {Celliers}, \citenamefont {Collins}, \citenamefont {Eggert},\ and\ \citenamefont {Jeanloz}}]{Millot2015}%
  \BibitemOpen
  \bibfield  {author} {\bibinfo {author} {\bibfnamefont {M.}~\bibnamefont {Millot}}, \bibinfo {author} {\bibfnamefont {N.}~\bibnamefont {Dubrovinskaia}}, \bibinfo {author} {\bibfnamefont {A.}~\bibnamefont {\u{C}ernok}}, \bibinfo {author} {\bibfnamefont {S.}~\bibnamefont {Blaha}}, \bibinfo {author} {\bibfnamefont {L.}~\bibnamefont {Dubrovinsky}}, \bibinfo {author} {\bibfnamefont {D.~G.}\ \bibnamefont {Braun}}, \bibinfo {author} {\bibfnamefont {P.~M.}\ \bibnamefont {Celliers}}, \bibinfo {author} {\bibfnamefont {G.~W.}\ \bibnamefont {Collins}}, \bibinfo {author} {\bibfnamefont {J.~H.}\ \bibnamefont {Eggert}},\ and\ \bibinfo {author} {\bibfnamefont {R.}~\bibnamefont {Jeanloz}},\ }\bibfield  {title} {\bibinfo {title} {Shock compression of stishovite and melting of silica at planetary interior conditions},\ }\href {https://doi.org/10.1126/science.1261507} {\bibfield  {journal} {\bibinfo  {journal} {Science}\ }\textbf {\bibinfo {volume} {347}},\ \bibinfo {pages} {418} (\bibinfo {year} {2015})}\BibitemShut {NoStop}%
\bibitem [{\citenamefont {Gonz\'{a}lez-Cataldo}\ \emph {et~al.}(2016{\natexlab{a}})\citenamefont {Gonz\'{a}lez-Cataldo}, \citenamefont {Davis},\ and\ \citenamefont {Guti\'{e}rrez}}]{Gonzales2016a}%
  \BibitemOpen
  \bibfield  {author} {\bibinfo {author} {\bibfnamefont {F.}~\bibnamefont {Gonz\'{a}lez-Cataldo}}, \bibinfo {author} {\bibfnamefont {S.}~\bibnamefont {Davis}},\ and\ \bibinfo {author} {\bibfnamefont {G.}~\bibnamefont {Guti\'{e}rrez}},\ }\bibfield  {title} {\bibinfo {title} {Melting curve of \text{SiO$_2$} at multimegabar pressures: implications for gas giants and super-earths},\ }\href {https://doi.org/10.1038/srep26537} {\bibfield  {journal} {\bibinfo  {journal} {Scientific Reports}\ }\textbf {\bibinfo {volume} {6}},\ \bibinfo {pages} {26537} (\bibinfo {year} {2016}{\natexlab{a}})}\BibitemShut {NoStop}%
\bibitem [{\citenamefont {Gonz\'{a}lez-Cataldo}\ \emph {et~al.}(2016{\natexlab{b}})\citenamefont {Gonz\'{a}lez-Cataldo}, \citenamefont {Davis},\ and\ \citenamefont {Guti\'{e}rrez}}]{Gonzalez2016b}%
  \BibitemOpen
  \bibfield  {author} {\bibinfo {author} {\bibfnamefont {F.}~\bibnamefont {Gonz\'{a}lez-Cataldo}}, \bibinfo {author} {\bibfnamefont {S.}~\bibnamefont {Davis}},\ and\ \bibinfo {author} {\bibfnamefont {G.}~\bibnamefont {Guti\'{e}rrez}},\ }\bibfield  {title} {\bibinfo {title} {Z method calculations to determine the melting curve of silica at high pressures},\ }\href {https://doi.org/10.1088/1742-6596/720/1/012032} {\bibfield  {journal} {\bibinfo  {journal} {Journal of Physics: Conference Series}\ }\textbf {\bibinfo {volume} {720}},\ \bibinfo {pages} {012032} (\bibinfo {year} {2016}{\natexlab{b}})}\BibitemShut {NoStop}%
\bibitem [{\citenamefont {Boates}\ and\ \citenamefont {Bonev}(2013)}]{Boates2013}%
  \BibitemOpen
  \bibfield  {author} {\bibinfo {author} {\bibfnamefont {B.}~\bibnamefont {Boates}}\ and\ \bibinfo {author} {\bibfnamefont {S.~A.}\ \bibnamefont {Bonev}},\ }\bibfield  {title} {\bibinfo {title} {Demixing instability in dense molten ${\mathrm{mgsio}}_{3}$ and the phase diagram of \text{MgO}},\ }\href {https://doi.org/10.1103/PhysRevLett.110.135504} {\bibfield  {journal} {\bibinfo  {journal} {Phys. Rev. Lett.}\ }\textbf {\bibinfo {volume} {110}},\ \bibinfo {pages} {135504} (\bibinfo {year} {2013})}\BibitemShut {NoStop}%
\bibitem [{\citenamefont {Hirose}(2017)}]{Hirose2017}%
  \BibitemOpen
  \bibfield  {author} {\bibinfo {author} {\bibfnamefont {K.}~\bibnamefont {Hirose}},\ }\bibfield  {title} {\bibinfo {title} {Crystallization of silicon dioxide and compositional evolution of the earth's core},\ }\href {https://doi.org/10.1038/nature21367} {\bibfield  {journal} {\bibinfo  {journal} {Nature}\ }\textbf {\bibinfo {volume} {543}},\ \bibinfo {pages} {99} (\bibinfo {year} {2017})}\BibitemShut {NoStop}%
\bibitem [{\citenamefont {Tr{\o}nnes}\ \emph {et~al.}(2019)\citenamefont {Tr{\o}nnes}, \citenamefont {Baron}, \citenamefont {Eigenmann}, \citenamefont {Guren}, \citenamefont {Heyn}, \citenamefont {L{\o}ken},\ and\ \citenamefont {Mohn}}]{Tronnes2019}%
  \BibitemOpen
  \bibfield  {author} {\bibinfo {author} {\bibfnamefont {R.~G.}\ \bibnamefont {Tr{\o}nnes}}, \bibinfo {author} {\bibfnamefont {M.~A.}\ \bibnamefont {Baron}}, \bibinfo {author} {\bibfnamefont {K.~R.}\ \bibnamefont {Eigenmann}}, \bibinfo {author} {\bibfnamefont {M.~G.}\ \bibnamefont {Guren}}, \bibinfo {author} {\bibfnamefont {B.~H.}\ \bibnamefont {Heyn}}, \bibinfo {author} {\bibfnamefont {A.}~\bibnamefont {L{\o}ken}},\ and\ \bibinfo {author} {\bibfnamefont {C.~E.}\ \bibnamefont {Mohn}},\ }\bibfield  {title} {\bibinfo {title} {Core formation, mantle differentiation and core-mantle interaction within earth and the terrestrial planets},\ }\href {https://doi.org/10.1016/j.tecto.2018.10.021} {\bibfield  {journal} {\bibinfo  {journal} {Tectonophysics}\ }\textbf {\bibinfo {volume} {760}},\ \bibinfo {pages} {165} (\bibinfo {year} {2019})}\BibitemShut {NoStop}%
\bibitem [{\citenamefont {Andrault}\ \emph {et~al.}(2022)\citenamefont {Andrault}, \citenamefont {Pison}, \citenamefont {Morard}, \citenamefont {Garbarino}, \citenamefont {Mezouar}, \citenamefont {Bouhifd},\ and\ \citenamefont {Kawamoto}}]{Andrault2022}%
  \BibitemOpen
  \bibfield  {author} {\bibinfo {author} {\bibfnamefont {D.}~\bibnamefont {Andrault}}, \bibinfo {author} {\bibfnamefont {L.}~\bibnamefont {Pison}}, \bibinfo {author} {\bibfnamefont {G.}~\bibnamefont {Morard}}, \bibinfo {author} {\bibfnamefont {G.}~\bibnamefont {Garbarino}}, \bibinfo {author} {\bibfnamefont {M.}~\bibnamefont {Mezouar}}, \bibinfo {author} {\bibfnamefont {M.~A.}\ \bibnamefont {Bouhifd}},\ and\ \bibinfo {author} {\bibfnamefont {T.}~\bibnamefont {Kawamoto}},\ }\bibfield  {title} {\bibinfo {title} {Comment on: Melting behavior of \text{SiO$_2$} up to 120 \text{GPa} (andrault et al. 2020)},\ }\href {https://doi.org/10.1007/s00269-021-01174-2} {\bibfield  {journal} {\bibinfo  {journal} {Physics and Chemistry of Minerals}\ }\textbf {\bibinfo {volume} {49}},\ \bibinfo {pages} {3} (\bibinfo {year} {2022})}\BibitemShut {NoStop}%
\bibitem [{\citenamefont {Andrault}\ \emph {et~al.}(2020)\citenamefont {Andrault}, \citenamefont {Morard}, \citenamefont {Garbarino}, \citenamefont {Mezouar}, \citenamefont {Bouhifd},\ and\ \citenamefont {Kawamoto}}]{Andrault2020}%
  \BibitemOpen
  \bibfield  {author} {\bibinfo {author} {\bibfnamefont {D.}~\bibnamefont {Andrault}}, \bibinfo {author} {\bibfnamefont {G.}~\bibnamefont {Morard}}, \bibinfo {author} {\bibfnamefont {G.}~\bibnamefont {Garbarino}}, \bibinfo {author} {\bibfnamefont {M.}~\bibnamefont {Mezouar}}, \bibinfo {author} {\bibfnamefont {M.~A.}\ \bibnamefont {Bouhifd}},\ and\ \bibinfo {author} {\bibfnamefont {T.}~\bibnamefont {Kawamoto}},\ }\bibfield  {title} {\bibinfo {title} {Melting behavior of \text{SiO$_2$} up to 120 \text{GPa}},\ }\href {https://doi.org/10.1007/s00269-019-01077-3} {\bibfield  {journal} {\bibinfo  {journal} {Physics and Chemistry of Minerals}\ }\textbf {\bibinfo {volume} {47}},\ \bibinfo {pages} {10} (\bibinfo {year} {2020})}\BibitemShut {NoStop}%
\bibitem [{\citenamefont {Belonoshko}\ and\ \citenamefont {Dubrovinsky}(1995)}]{Belonoshko1995}%
  \BibitemOpen
  \bibfield  {author} {\bibinfo {author} {\bibfnamefont {A.~B.}\ \bibnamefont {Belonoshko}}\ and\ \bibinfo {author} {\bibfnamefont {L.~S.}\ \bibnamefont {Dubrovinsky}},\ }\bibfield  {title} {\bibinfo {title} {Molecular dynamics of stishovite melting},\ }\href {https://doi.org/10.1016/0016-7037(95)00071-7} {\bibfield  {journal} {\bibinfo  {journal} {Geochimica et Cosmochimica Acta}\ }\textbf {\bibinfo {volume} {59}},\ \bibinfo {pages} {1883} (\bibinfo {year} {1995})}\BibitemShut {NoStop}%
\bibitem [{\citenamefont {Akaogi}\ \emph {et~al.}(2011)\citenamefont {Akaogi}, \citenamefont {Oohata}, \citenamefont {Kojitani},\ and\ \citenamefont {Kawaji}}]{Akaogi2011}%
  \BibitemOpen
  \bibfield  {author} {\bibinfo {author} {\bibfnamefont {M.}~\bibnamefont {Akaogi}}, \bibinfo {author} {\bibfnamefont {M.}~\bibnamefont {Oohata}}, \bibinfo {author} {\bibfnamefont {H.}~\bibnamefont {Kojitani}},\ and\ \bibinfo {author} {\bibfnamefont {H.}~\bibnamefont {Kawaji}},\ }\bibfield  {title} {\bibinfo {title} {Thermodynamic properties of stishovite by low-temperature heat capacity measurements and the coesite-stishovite transition boundary},\ }\href {https://doi.org/10.2138/am.2011.3748} {\bibfield  {journal} {\bibinfo  {journal} {American Mineralogist}\ }\textbf {\bibinfo {volume} {96}},\ \bibinfo {pages} {1325} (\bibinfo {year} {2011})}\BibitemShut {NoStop}%
\bibitem [{\citenamefont {Kuwayama}\ \emph {et~al.}(2005)\citenamefont {Kuwayama}, \citenamefont {Hirose}, \citenamefont {Sata},\ and\ \citenamefont {Ohishi}}]{Kuwayama2005}%
  \BibitemOpen
  \bibfield  {author} {\bibinfo {author} {\bibfnamefont {Y.}~\bibnamefont {Kuwayama}}, \bibinfo {author} {\bibfnamefont {K.}~\bibnamefont {Hirose}}, \bibinfo {author} {\bibfnamefont {N.}~\bibnamefont {Sata}},\ and\ \bibinfo {author} {\bibfnamefont {Y.}~\bibnamefont {Ohishi}},\ }\bibfield  {title} {\bibinfo {title} {The pyrite-type high-pressure form of silica},\ }\href {https://doi.org/10.1126/science.1114879} {\bibfield  {journal} {\bibinfo  {journal} {Science}\ }\textbf {\bibinfo {volume} {309}},\ \bibinfo {pages} {923} (\bibinfo {year} {2005})}\BibitemShut {NoStop}%
\bibitem [{\citenamefont {Liu}\ \emph {et~al.}(2021)\citenamefont {Liu}, \citenamefont {Shi}, \citenamefont {Gao}, \citenamefont {Wang}, \citenamefont {Han}, \citenamefont {Lu}, \citenamefont {Wang}, \citenamefont {Xing},\ and\ \citenamefont {Sun}}]{SolidCN2021}%
  \BibitemOpen
  \bibfield  {author} {\bibinfo {author} {\bibfnamefont {C.}~\bibnamefont {Liu}}, \bibinfo {author} {\bibfnamefont {J.}~\bibnamefont {Shi}}, \bibinfo {author} {\bibfnamefont {H.}~\bibnamefont {Gao}}, \bibinfo {author} {\bibfnamefont {J.}~\bibnamefont {Wang}}, \bibinfo {author} {\bibfnamefont {Y.}~\bibnamefont {Han}}, \bibinfo {author} {\bibfnamefont {X.}~\bibnamefont {Lu}}, \bibinfo {author} {\bibfnamefont {H.-T.}\ \bibnamefont {Wang}}, \bibinfo {author} {\bibfnamefont {D.}~\bibnamefont {Xing}},\ and\ \bibinfo {author} {\bibfnamefont {J.}~\bibnamefont {Sun}},\ }\bibfield  {title} {\bibinfo {title} {Mixed coordination silica at megabar pressure},\ }\href {https://doi.org/10.1103/PhysRevLett.126.035701} {\bibfield  {journal} {\bibinfo  {journal} {Physical Review Letters}\ }\textbf {\bibinfo {volume} {126}},\ \bibinfo {pages} {035701} (\bibinfo {year} {2021})}\BibitemShut {NoStop}%
\bibitem [{\citenamefont {Li}\ \emph {et~al.}(2023)\citenamefont {Li}, \citenamefont {Zhang}, \citenamefont {Niu}, \citenamefont {Liu}, \citenamefont {Li}, \citenamefont {Wang},\ and\ \citenamefont {Zhang}}]{Li2023}%
  \BibitemOpen
  \bibfield  {author} {\bibinfo {author} {\bibfnamefont {G.}~\bibnamefont {Li}}, \bibinfo {author} {\bibfnamefont {Z.}~\bibnamefont {Zhang}}, \bibinfo {author} {\bibfnamefont {X.}~\bibnamefont {Niu}}, \bibinfo {author} {\bibfnamefont {J.}~\bibnamefont {Liu}}, \bibinfo {author} {\bibfnamefont {J.}~\bibnamefont {Li}}, \bibinfo {author} {\bibfnamefont {W.}~\bibnamefont {Wang}},\ and\ \bibinfo {author} {\bibfnamefont {Z.}~\bibnamefont {Zhang}},\ }\bibfield  {title} {\bibinfo {title} {Equations of state and phase boundaries of \text{SiO$_2$} polymorphs under lower mantle conditions},\ }\href {https://doi.org/10.1029/2023JB026774} {\bibfield  {journal} {\bibinfo  {journal} {Journal of Geophysical Research: Solid Earth}\ }\textbf {\bibinfo {volume} {128}},\ \bibinfo {pages} {e2023JB026774} (\bibinfo {year} {2023})}\BibitemShut {NoStop}%
\bibitem [{\citenamefont {Hong}\ and\ \citenamefont {{van de Walle}}(2016)}]{Hong2016}%
  \BibitemOpen
  \bibfield  {author} {\bibinfo {author} {\bibfnamefont {Q.-J.}\ \bibnamefont {Hong}}\ and\ \bibinfo {author} {\bibfnamefont {A.}~\bibnamefont {{van de Walle}}},\ }\bibfield  {title} {\bibinfo {title} {A user guide for sluschi: Solid and liquid in ultra small coexistence with hovering interfaces},\ }\href {https://doi.org/10.1016/j.calphad.2015.12.003} {\bibfield  {journal} {\bibinfo  {journal} {Calphad}\ }\textbf {\bibinfo {volume} {52}},\ \bibinfo {pages} {88} (\bibinfo {year} {2016})}\BibitemShut {NoStop}%
\bibitem [{\citenamefont {Hernandez}\ \emph {et~al.}(2022)\citenamefont {Hernandez}, \citenamefont {Mohn}, \citenamefont {Guren}, \citenamefont {Baron},\ and\ \citenamefont {Tr{\o}nnes}}]{Hernandez2022}%
  \BibitemOpen
  \bibfield  {author} {\bibinfo {author} {\bibfnamefont {J.-A.}\ \bibnamefont {Hernandez}}, \bibinfo {author} {\bibfnamefont {C.~E.}\ \bibnamefont {Mohn}}, \bibinfo {author} {\bibfnamefont {M.~G.}\ \bibnamefont {Guren}}, \bibinfo {author} {\bibfnamefont {M.~A.}\ \bibnamefont {Baron}},\ and\ \bibinfo {author} {\bibfnamefont {R.~G.}\ \bibnamefont {Tr{\o}nnes}},\ }\bibfield  {title} {\bibinfo {title} {Ab initio atomistic simulations of \text{Ca}-perovskite melting},\ }\href {https://doi.org/10.1029/2021GL097262} {\bibfield  {journal} {\bibinfo  {journal} {Geophysical Research Letters}\ }\textbf {\bibinfo {volume} {49}},\ \bibinfo {pages} {e2021GL097262} (\bibinfo {year} {2022})}\BibitemShut {NoStop}%
\bibitem [{\citenamefont {Dorner}\ \emph {et~al.}(2018)\citenamefont {Dorner}, \citenamefont {Sukurma}, \citenamefont {Dellago},\ and\ \citenamefont {Kresse}}]{TI2018}%
  \BibitemOpen
  \bibfield  {author} {\bibinfo {author} {\bibfnamefont {F.}~\bibnamefont {Dorner}}, \bibinfo {author} {\bibfnamefont {Z.}~\bibnamefont {Sukurma}}, \bibinfo {author} {\bibfnamefont {C.}~\bibnamefont {Dellago}},\ and\ \bibinfo {author} {\bibfnamefont {G.}~\bibnamefont {Kresse}},\ }\bibfield  {title} {\bibinfo {title} {Melting si: Beyond density functional theory},\ }\href {https://doi.org/10.1103/PhysRevLett.121.195701} {\bibfield  {journal} {\bibinfo  {journal} {Physical Review Letters}\ }\textbf {\bibinfo {volume} {121}},\ \bibinfo {pages} {195701} (\bibinfo {year} {2018})}\BibitemShut {NoStop}%
\bibitem [{\citenamefont {Sun}\ \emph {et~al.}(2018)\citenamefont {Sun}, \citenamefont {Brodholt}, \citenamefont {Li},\ and\ \citenamefont {Vo\v{c}adlo}}]{Melting2018}%
  \BibitemOpen
  \bibfield  {author} {\bibinfo {author} {\bibfnamefont {T.}~\bibnamefont {Sun}}, \bibinfo {author} {\bibfnamefont {J.~P.}\ \bibnamefont {Brodholt}}, \bibinfo {author} {\bibfnamefont {Y.}~\bibnamefont {Li}},\ and\ \bibinfo {author} {\bibfnamefont {L.}~\bibnamefont {Vo\v{c}adlo}},\ }\bibfield  {title} {\bibinfo {title} {Melting properties from ab initio free energy calculations: Iron at the earth's inner-core boundary},\ }\href {https://doi.org/10.1103/PhysRevB.98.224301} {\bibfield  {journal} {\bibinfo  {journal} {Physical Review B}\ }\textbf {\bibinfo {volume} {98}},\ \bibinfo {pages} {224301} (\bibinfo {year} {2018})}\BibitemShut {NoStop}%
\bibitem [{\citenamefont {Rang}\ and\ \citenamefont {Kresse}(2019)}]{TI2019}%
  \BibitemOpen
  \bibfield  {author} {\bibinfo {author} {\bibfnamefont {M.}~\bibnamefont {Rang}}\ and\ \bibinfo {author} {\bibfnamefont {G.}~\bibnamefont {Kresse}},\ }\bibfield  {title} {\bibinfo {title} {First-principles study of the melting temperature of \text{MgO}},\ }\href {https://doi.org/10.1103/PhysRevB.99.184103} {\bibfield  {journal} {\bibinfo  {journal} {Physical Review B}\ }\textbf {\bibinfo {volume} {99}},\ \bibinfo {pages} {184103} (\bibinfo {year} {2019})}\BibitemShut {NoStop}%
\bibitem [{\citenamefont {Togo}\ and\ \citenamefont {Tanaka}(2015)}]{Phononpy2015}%
  \BibitemOpen
  \bibfield  {author} {\bibinfo {author} {\bibfnamefont {A.}~\bibnamefont {Togo}}\ and\ \bibinfo {author} {\bibfnamefont {I.}~\bibnamefont {Tanaka}},\ }\bibfield  {title} {\bibinfo {title} {First principles phonon calculations in materials science},\ }\href {https://doi.org/10.1016/j.scriptamat.2015.07.021} {\bibfield  {journal} {\bibinfo  {journal} {Scripta Materialia}\ }\textbf {\bibinfo {volume} {108}},\ \bibinfo {pages} {1} (\bibinfo {year} {2015})}\BibitemShut {NoStop}%
\bibitem [{\citenamefont {Belonoshko}\ \emph {et~al.}(2006)\citenamefont {Belonoshko}, \citenamefont {Skorodumova}, \citenamefont {Rosengren},\ and\ \citenamefont {Johansson}}]{Belonoshko2006}%
  \BibitemOpen
  \bibfield  {author} {\bibinfo {author} {\bibfnamefont {A.~B.}\ \bibnamefont {Belonoshko}}, \bibinfo {author} {\bibfnamefont {N.~V.}\ \bibnamefont {Skorodumova}}, \bibinfo {author} {\bibfnamefont {A.}~\bibnamefont {Rosengren}},\ and\ \bibinfo {author} {\bibfnamefont {B.}~\bibnamefont {Johansson}},\ }\bibfield  {title} {\bibinfo {title} {Melting and critical superheating},\ }\href {https://doi.org/10.1103/PhysRevB.73.012201} {\bibfield  {journal} {\bibinfo  {journal} {Physical Review B}\ }\textbf {\bibinfo {volume} {73}},\ \bibinfo {pages} {012201} (\bibinfo {year} {2006})}\BibitemShut {NoStop}%
\bibitem [{\citenamefont {Alf\'e}\ \emph {et~al.}(2011)\citenamefont {Alf\'e}, \citenamefont {Cazorla},\ and\ \citenamefont {Gillan}}]{Alfe2011}%
  \BibitemOpen
  \bibfield  {author} {\bibinfo {author} {\bibfnamefont {D.}~\bibnamefont {Alf\'e}}, \bibinfo {author} {\bibfnamefont {C.}~\bibnamefont {Cazorla}},\ and\ \bibinfo {author} {\bibfnamefont {M.~J.}\ \bibnamefont {Gillan}},\ }\bibfield  {title} {\bibinfo {title} {The kinetics of homogeneous melting beyond the limit of superheating},\ }\href {https://doi.org/10.1063/1.3605601} {\bibfield  {journal} {\bibinfo  {journal} {The Journal of Chemical Physics}\ }\textbf {\bibinfo {volume} {135}},\ \bibinfo {pages} {024102} (\bibinfo {year} {2011})}\BibitemShut {NoStop}%
\bibitem [{\citenamefont {Perdew}\ \emph {et~al.}(1996)\citenamefont {Perdew}, \citenamefont {Burke},\ and\ \citenamefont {Ernzerhof}}]{PBE96}%
  \BibitemOpen
  \bibfield  {author} {\bibinfo {author} {\bibfnamefont {J.~P.}\ \bibnamefont {Perdew}}, \bibinfo {author} {\bibfnamefont {K.}~\bibnamefont {Burke}},\ and\ \bibinfo {author} {\bibfnamefont {M.}~\bibnamefont {Ernzerhof}},\ }\bibfield  {title} {\bibinfo {title} {Generalized gradient approximation made simple},\ }\href {https://doi.org/10.1103/PhysRevLett.77.3865} {\bibfield  {journal} {\bibinfo  {journal} {Physical Review Letters}\ }\textbf {\bibinfo {volume} {77}},\ \bibinfo {pages} {3865} (\bibinfo {year} {1996})}\BibitemShut {NoStop}%
\bibitem [{\citenamefont {Braithwaite}\ and\ \citenamefont {Stixrude}(2019)}]{Braithwaite2019}%
  \BibitemOpen
  \bibfield  {author} {\bibinfo {author} {\bibfnamefont {J.}~\bibnamefont {Braithwaite}}\ and\ \bibinfo {author} {\bibfnamefont {L.}~\bibnamefont {Stixrude}},\ }\bibfield  {title} {\bibinfo {title} {Melting of \text{CaSiO$_3$} perovskite at high pressure},\ }\href {https://doi.org/10.1029/2018GL081805} {\bibfield  {journal} {\bibinfo  {journal} {Geophysical Research Letters}\ }\textbf {\bibinfo {volume} {46}},\ \bibinfo {pages} {2037} (\bibinfo {year} {2019})}\BibitemShut {NoStop}%
\bibitem [{\citenamefont {Mermin}(1965)}]{Mermin1965}%
  \BibitemOpen
  \bibfield  {author} {\bibinfo {author} {\bibfnamefont {N.~D.}\ \bibnamefont {Mermin}},\ }\bibfield  {title} {\bibinfo {title} {Thermal properties of the inhomogeneous electron gas},\ }\href {https://doi.org/10.1103/PhysRev.137.A1441} {\bibfield  {journal} {\bibinfo  {journal} {Physical Review}\ }\textbf {\bibinfo {volume} {137}},\ \bibinfo {pages} {A1441} (\bibinfo {year} {1965})}\BibitemShut {NoStop}%
\bibitem [{\citenamefont {Wentzcovitch}\ \emph {et~al.}(1992)\citenamefont {Wentzcovitch}, \citenamefont {Martins},\ and\ \citenamefont {Allen}}]{Wentzcovitch1992}%
  \BibitemOpen
  \bibfield  {author} {\bibinfo {author} {\bibfnamefont {R.~M.}\ \bibnamefont {Wentzcovitch}}, \bibinfo {author} {\bibfnamefont {J.~L.}\ \bibnamefont {Martins}},\ and\ \bibinfo {author} {\bibfnamefont {P.~B.}\ \bibnamefont {Allen}},\ }\bibfield  {title} {\bibinfo {title} {Energy versus free-energy conservation in first-principles molecular dynamics},\ }\href {https://doi.org/10.1103/PhysRevB.45.11372} {\bibfield  {journal} {\bibinfo  {journal} {Physical Review B}\ }\textbf {\bibinfo {volume} {45}},\ \bibinfo {pages} {11372} (\bibinfo {year} {1992})}\BibitemShut {NoStop}%
\bibitem [{\citenamefont {Geng}\ and\ \citenamefont {Mohn}(2023)}]{ZMethod2023}%
  \BibitemOpen
  \bibfield  {author} {\bibinfo {author} {\bibfnamefont {M.}~\bibnamefont {Geng}}\ and\ \bibinfo {author} {\bibfnamefont {C.~E.}\ \bibnamefont {Mohn}},\ }\bibfield  {title} {\bibinfo {title} {Influence of electronic entropy on hellmann-feynman forces in ab initio molecular dynamics with large temperature changes},\ }\href {https://doi.org/10.1103/PhysRevB.108.134110} {\bibfield  {journal} {\bibinfo  {journal} {Physical Review B}\ }\textbf {\bibinfo {volume} {108}},\ \bibinfo {pages} {134110} (\bibinfo {year} {2023})}\BibitemShut {NoStop}%
\bibitem [{\citenamefont {Kresse}\ and\ \citenamefont {Hafner}(1993)}]{VASP93}%
  \BibitemOpen
  \bibfield  {author} {\bibinfo {author} {\bibfnamefont {G.}~\bibnamefont {Kresse}}\ and\ \bibinfo {author} {\bibfnamefont {J.}~\bibnamefont {Hafner}},\ }\bibfield  {title} {\bibinfo {title} {Ab initio molecular dynamics for liquid metals},\ }\href {https://doi.org/10.1103/PhysRevB.47.558} {\bibfield  {journal} {\bibinfo  {journal} {Physical Review B}\ }\textbf {\bibinfo {volume} {47}},\ \bibinfo {pages} {558} (\bibinfo {year} {1993})}\BibitemShut {NoStop}%
\bibitem [{\citenamefont {Kresse}\ and\ \citenamefont {Hafner}(1994)}]{VASP94}%
  \BibitemOpen
  \bibfield  {author} {\bibinfo {author} {\bibfnamefont {G.}~\bibnamefont {Kresse}}\ and\ \bibinfo {author} {\bibfnamefont {J.}~\bibnamefont {Hafner}},\ }\bibfield  {title} {\bibinfo {title} {Ab initio molecular-dynamics simulation of the liquid-metal-amorphous-semiconductor transition in germanium},\ }\href {https://doi.org/10.1103/PhysRevB.49.14251} {\bibfield  {journal} {\bibinfo  {journal} {Physical Review B}\ }\textbf {\bibinfo {volume} {49}},\ \bibinfo {pages} {14251} (\bibinfo {year} {1994})}\BibitemShut {NoStop}%
\bibitem [{\citenamefont {Bl\"ochl}(1994)}]{PAW94}%
  \BibitemOpen
  \bibfield  {author} {\bibinfo {author} {\bibfnamefont {P.~E.}\ \bibnamefont {Bl\"ochl}},\ }\bibfield  {title} {\bibinfo {title} {Projector augmented-wave method},\ }\href {https://doi.org/10.1103/PhysRevB.50.17953} {\bibfield  {journal} {\bibinfo  {journal} {Physical Review B}\ }\textbf {\bibinfo {volume} {50}},\ \bibinfo {pages} {17953} (\bibinfo {year} {1994})}\BibitemShut {NoStop}%
\bibitem [{\citenamefont {Kresse}\ and\ \citenamefont {Joubert}(1999)}]{PAW99}%
  \BibitemOpen
  \bibfield  {author} {\bibinfo {author} {\bibfnamefont {G.}~\bibnamefont {Kresse}}\ and\ \bibinfo {author} {\bibfnamefont {D.}~\bibnamefont {Joubert}},\ }\bibfield  {title} {\bibinfo {title} {From ultrasoft pseudopotentials to the projector augmented-wave method},\ }\href {https://doi.org/10.1103/PhysRevB.59.1758} {\bibfield  {journal} {\bibinfo  {journal} {Physical Review B}\ }\textbf {\bibinfo {volume} {59}},\ \bibinfo {pages} {1758} (\bibinfo {year} {1999})}\BibitemShut {NoStop}%
\bibitem [{\citenamefont {Akimoto}\ and\ \citenamefont {Syono}(1969)}]{CoeTrans1969}%
  \BibitemOpen
  \bibfield  {author} {\bibinfo {author} {\bibfnamefont {S.-i.}\ \bibnamefont {Akimoto}}\ and\ \bibinfo {author} {\bibfnamefont {Y.}~\bibnamefont {Syono}},\ }\bibfield  {title} {\bibinfo {title} {Coesite-stishovite transition},\ }\href {https://doi.org/10.1029/JB074i006p01653} {\bibfield  {journal} {\bibinfo  {journal} {Journal of Geophysical Research (1896-1977)}\ }\textbf {\bibinfo {volume} {74}},\ \bibinfo {pages} {1653} (\bibinfo {year} {1969})}\BibitemShut {NoStop}%
\bibitem [{\citenamefont {Yagi}\ and\ \citenamefont {Akimoto}(1976)}]{CoeTrans1976}%
  \BibitemOpen
  \bibfield  {author} {\bibinfo {author} {\bibfnamefont {T.}~\bibnamefont {Yagi}}\ and\ \bibinfo {author} {\bibfnamefont {S.-I.}\ \bibnamefont {Akimoto}},\ }\bibfield  {title} {\bibinfo {title} {Direct determination of coesite- stishovite transition by in-situ x-ray measurements},\ }\href {https://doi.org/10.1016/0040-1951(76)90042-1} {\bibfield  {journal} {\bibinfo  {journal} {Tectonophysics}\ }\textbf {\bibinfo {volume} {35}},\ \bibinfo {pages} {259} (\bibinfo {year} {1976})}\BibitemShut {NoStop}%
\bibitem [{\citenamefont {Serghiou}\ \emph {et~al.}(1995)\citenamefont {Serghiou}, \citenamefont {Zerr}, \citenamefont {Chudinovskikh},\ and\ \citenamefont {Boehler}}]{CoeTrans1995}%
  \BibitemOpen
  \bibfield  {author} {\bibinfo {author} {\bibfnamefont {G.}~\bibnamefont {Serghiou}}, \bibinfo {author} {\bibfnamefont {A.}~\bibnamefont {Zerr}}, \bibinfo {author} {\bibfnamefont {L.}~\bibnamefont {Chudinovskikh}},\ and\ \bibinfo {author} {\bibfnamefont {R.}~\bibnamefont {Boehler}},\ }\bibfield  {title} {\bibinfo {title} {The coesite-stishovite transition in a laser-heated diamond cell},\ }\href {https://doi.org/10.1029/94GL02692} {\bibfield  {journal} {\bibinfo  {journal} {Geophysical Research Letters}\ }\textbf {\bibinfo {volume} {22}},\ \bibinfo {pages} {441} (\bibinfo {year} {1995})}\BibitemShut {NoStop}%
\bibitem [{\citenamefont {Zhang}\ \emph {et~al.}(1996)\citenamefont {Zhang}, \citenamefont {Li}, \citenamefont {Utsumi},\ and\ \citenamefont {Liebermann}}]{CoeTrans1996}%
  \BibitemOpen
  \bibfield  {author} {\bibinfo {author} {\bibfnamefont {J.}~\bibnamefont {Zhang}}, \bibinfo {author} {\bibfnamefont {B.}~\bibnamefont {Li}}, \bibinfo {author} {\bibfnamefont {W.}~\bibnamefont {Utsumi}},\ and\ \bibinfo {author} {\bibfnamefont {R.~C.}\ \bibnamefont {Liebermann}},\ }\bibfield  {title} {\bibinfo {title} {In situ x-ray observations of the coesite-stishovite transition: reversed phase boundary and kinetics},\ }\href {https://doi.org/10.1007/BF00202987} {\bibfield  {journal} {\bibinfo  {journal} {Physics and Chemistry of Minerals}\ }\textbf {\bibinfo {volume} {23}},\ \bibinfo {pages} {1} (\bibinfo {year} {1996})}\BibitemShut {NoStop}%
\bibitem [{\citenamefont {Hu}\ \emph {et~al.}(2015)\citenamefont {Hu}, \citenamefont {Shu}, \citenamefont {Cadien}, \citenamefont {Meng}, \citenamefont {Yang}, \citenamefont {Sheng},\ and\ \citenamefont {Mao}}]{CoeTrans2015}%
  \BibitemOpen
  \bibfield  {author} {\bibinfo {author} {\bibfnamefont {Q.~Y.}\ \bibnamefont {Hu}}, \bibinfo {author} {\bibfnamefont {J.~F.}\ \bibnamefont {Shu}}, \bibinfo {author} {\bibfnamefont {A.}~\bibnamefont {Cadien}}, \bibinfo {author} {\bibfnamefont {Y.}~\bibnamefont {Meng}}, \bibinfo {author} {\bibfnamefont {W.~G.}\ \bibnamefont {Yang}}, \bibinfo {author} {\bibfnamefont {H.~W.}\ \bibnamefont {Sheng}},\ and\ \bibinfo {author} {\bibfnamefont {H.~K.}\ \bibnamefont {Mao}},\ }\bibfield  {title} {\bibinfo {title} {Polymorphic phase transition mechanism of compressed coesite},\ }\href {https://doi.org/10.1038/ncomms7630} {\bibfield  {journal} {\bibinfo  {journal} {Nature Communications}\ }\textbf {\bibinfo {volume} {6}},\ \bibinfo {pages} {6630} (\bibinfo {year} {2015})}\BibitemShut {NoStop}%
\bibitem [{\citenamefont {Krug}\ \emph {et~al.}(2022)\citenamefont {Krug}, \citenamefont {Saki}, \citenamefont {Ledoux}, \citenamefont {Gay}, \citenamefont {Chantel}, \citenamefont {Pakhomova}, \citenamefont {Husband}, \citenamefont {Rohrbach}, \citenamefont {Klemme}, \citenamefont {Thomas}, \citenamefont {Merkel},\ and\ \citenamefont {Sanchez-Valle}}]{CoeTrans2022}%
  \BibitemOpen
  \bibfield  {author} {\bibinfo {author} {\bibfnamefont {M.}~\bibnamefont {Krug}}, \bibinfo {author} {\bibfnamefont {M.}~\bibnamefont {Saki}}, \bibinfo {author} {\bibfnamefont {E.}~\bibnamefont {Ledoux}}, \bibinfo {author} {\bibfnamefont {J.~P.}\ \bibnamefont {Gay}}, \bibinfo {author} {\bibfnamefont {J.}~\bibnamefont {Chantel}}, \bibinfo {author} {\bibfnamefont {A.}~\bibnamefont {Pakhomova}}, \bibinfo {author} {\bibfnamefont {R.}~\bibnamefont {Husband}}, \bibinfo {author} {\bibfnamefont {A.}~\bibnamefont {Rohrbach}}, \bibinfo {author} {\bibfnamefont {S.}~\bibnamefont {Klemme}}, \bibinfo {author} {\bibfnamefont {C.}~\bibnamefont {Thomas}}, \bibinfo {author} {\bibfnamefont {S.}~\bibnamefont {Merkel}},\ and\ \bibinfo {author} {\bibfnamefont {C.}~\bibnamefont {Sanchez-Valle}},\ }\bibfield  {title} {\bibinfo {title} {Textures induced by the coesite-stishovite transition and implications for the visibility of the x-discontinuity},\ }\href {https://doi.org/10.1029/2022GC010544} {\bibfield  {journal} {\bibinfo  {journal} {Geochemistry, Geophysics, Geosystems}\ }\textbf {\bibinfo {volume} {23}},\ \bibinfo {pages} {e2022GC010544} (\bibinfo {year} {2022})}\BibitemShut {NoStop}%
\bibitem [{\citenamefont {Shen}\ \emph {et~al.}(2016)\citenamefont {Shen}, \citenamefont {Jester}, \citenamefont {Qi},\ and\ \citenamefont {Reed}}]{Shen2016}%
  \BibitemOpen
  \bibfield  {author} {\bibinfo {author} {\bibfnamefont {Y.}~\bibnamefont {Shen}}, \bibinfo {author} {\bibfnamefont {S.~B.}\ \bibnamefont {Jester}}, \bibinfo {author} {\bibfnamefont {T.}~\bibnamefont {Qi}},\ and\ \bibinfo {author} {\bibfnamefont {E.~J.}\ \bibnamefont {Reed}},\ }\bibfield  {title} {\bibinfo {title} {Nanosecond homogeneous nucleation and crystal growth in shock-compressed \text{SiO$_2$}},\ }\href {https://doi.org/10.1038/nmat4447} {\bibfield  {journal} {\bibinfo  {journal} {Nature Materials}\ }\textbf {\bibinfo {volume} {15}},\ \bibinfo {pages} {60} (\bibinfo {year} {2016})}\BibitemShut {NoStop}%
\bibitem [{\citenamefont {Niu}\ \emph {et~al.}(2018)\citenamefont {Niu}, \citenamefont {Piaggi}, \citenamefont {Invernizzi},\ and\ \citenamefont {Parrinello}}]{liqsilica2018}%
  \BibitemOpen
  \bibfield  {author} {\bibinfo {author} {\bibfnamefont {H.}~\bibnamefont {Niu}}, \bibinfo {author} {\bibfnamefont {P.~M.}\ \bibnamefont {Piaggi}}, \bibinfo {author} {\bibfnamefont {M.}~\bibnamefont {Invernizzi}},\ and\ \bibinfo {author} {\bibfnamefont {M.}~\bibnamefont {Parrinello}},\ }\bibfield  {title} {\bibinfo {title} {Molecular dynamics simulations of liquid silica crystallization},\ }\href {https://doi.org/10.1073/pnas.1803919115} {\bibfield  {journal} {\bibinfo  {journal} {Proceedings of the National Academy of Sciences}\ }\textbf {\bibinfo {volume} {115}},\ \bibinfo {pages} {5348} (\bibinfo {year} {2018})}\BibitemShut {NoStop}%
\bibitem [{\citenamefont {Kalkan}\ \emph {et~al.}(2022)\citenamefont {Kalkan}, \citenamefont {Godwal}, \citenamefont {Yan},\ and\ \citenamefont {Jeanloz}}]{HPLiquid}%
  \BibitemOpen
  \bibfield  {author} {\bibinfo {author} {\bibfnamefont {B.}~\bibnamefont {Kalkan}}, \bibinfo {author} {\bibfnamefont {B.~K.}\ \bibnamefont {Godwal}}, \bibinfo {author} {\bibfnamefont {J.}~\bibnamefont {Yan}},\ and\ \bibinfo {author} {\bibfnamefont {R.}~\bibnamefont {Jeanloz}},\ }\bibfield  {title} {\bibinfo {title} {High-pressure phase transitions and melt structure of $\mathrm{PbO}_{2}$: An analog for silica},\ }\href {https://doi.org/10.1103/PhysRevB.105.064111} {\bibfield  {journal} {\bibinfo  {journal} {Physical Review B}\ }\textbf {\bibinfo {volume} {105}},\ \bibinfo {pages} {064111} (\bibinfo {year} {2022})}\BibitemShut {NoStop}%
\bibitem [{\citenamefont {Zimmermann}\ and\ \citenamefont {Jain}(2020)}]{CrystalNN2020}%
  \BibitemOpen
  \bibfield  {author} {\bibinfo {author} {\bibfnamefont {N.~E.~R.}\ \bibnamefont {Zimmermann}}\ and\ \bibinfo {author} {\bibfnamefont {A.}~\bibnamefont {Jain}},\ }\bibfield  {title} {\bibinfo {title} {Local structure order parameters and site fingerprints for quantification of coordination environment and crystal structure similarity},\ }\href {https://doi.org/10.1039/C9RA07755C} {\bibfield  {journal} {\bibinfo  {journal} {RSC Advances}\ }\textbf {\bibinfo {volume} {10}},\ \bibinfo {pages} {6063} (\bibinfo {year} {2020})}\BibitemShut {NoStop}%
\end{thebibliography}%

\end{document}